%% file: main.tex
\documentclass[acmsmall,screen,nonacm,natbib=false]{acmart}

\AtBeginDocument{%
  }

\RequirePackage[
  datamodel=acmdatamodel,
  style=numeric,
  ]{biblatex}

\input{preamble}

\begin{document}

\title{The Fearless Journey}

\author{Nick Webster}
\email{nick.webster@vuw.ac.nz}
\orcid{0000-0002-2508-6643}
\author{Marco Servetto}
\email{marco.servetto@ecs.vuw.ac.nz}
\orcid{0000-0003-1458-2868}
\author{Michael Homer}
\email{michael.homer@ecs.vuw.ac.nz}
\orcid{0000-0003-0280-6748}
\affiliation{%
  \institution{Victoria University of Wellington}
  \streetaddress{P.O. Box 600}
  \city{Wellington}
  \country{New Zealand}
  \postcode{6140}
}

\renewcommand{\shortauthors}{Webster, Servetto, and Homer}

\begin{abstract}
Existing minimal Object-Oriented models (OO), like Featherweight Java (FJ), are valuable for modelling programs and designing new programming languages and tools. 
However, their utility in developing real-world programs is limited. We introduce the `Fearless Heart', a novel object calculus preserving FJ's minimal and extensible nature while being more suited for constructing complex, real-world applications.

To illustrate the extensibility of the Fearless Heart, we extend it with Reference Capabilities (RC), creating R-Fearless. It supports mutability and other side effects while retaining the reasoning advantages of functional programming 
and gaining support for features that are well-known to be enabled by RC,
like automatic parallelism, caching and invariants.
R-Fearless is still minimal enough to allow further extensions. It is an ideal foundation for constructing both practical systems and formal models.
\end{abstract}

%\ccsdesc[300]{Theory of computation~Invariants}
%commented to save space, can be ripristinated later



\maketitle

\section{Introduction}
It is hard to overstate the beneficial impact of Featherweight Java (FJ) \cite{fj} on the world of formal PL modelling. Many works have extended FJ over the years \cite{bierman2003mj,Murawski2014,toplas42,10.1145/3611096.3611098}, showing the value of a powerful core language.
When using FJ as a base for extensions and presenting examples using our extensions, we often assume to have even more features, like primitive types, statements, and local variables. However, this can be brittle: those other features are usually outside of the formal model in the paper, not included in proofs, and thus may be a source of hidden unsoundness \cite{10.1145/2983990.2984004}.
%For example, consider someone building a new language by extending FJ and adding some dataflow analysis for secure programming \cite{toplas42}. The hand-wave introduction of features like if-statements or while-loops would introduce new forms of information leakage unaddressed by the original model.

% ------The Fearless Heart avoids this problem by making it trivial to encode user-friendly support for all of these features at the library level with relatively small amounts of code. Moving all of these language features to the library level means the Fearless Heart itself remains simple. Additionally, implementing a feature like booleans with if-else in the Fearless Heart is not just writing a program; it is a formal model of that language feature by virtue of it being written in the syntax and semantics of the Fearless Heart's formal model.

%In our pursuit of modelling Java subsets and building on years of experience we encountered challenges due to the complexity of the language features we had to contend with.  We sought ways 
% We are modelling Java subsets in order to simplify the language without losing expressive power.
We made some intriguing discoveries when modelling Java subsets to simplify the language without losing expressive power. We realised that \lstinline[language=Java]{static} could be replaced with singleton objects, akin to Scala's companion objects \cite{Alexander2021-rw}.
We could eliminate  constructors by introducing object literals, like anonymous inner classes and lambdas in Java. Finally, we found that fields, a fundamental element of object-oriented languages, could be substituted by capturing all state in closures. As we converged on this design, our model began to resemble lambda calculus, while still being practically usable as an actual programming language.
%, which we will discuss more in \autoref{sec:core}.
We call this new minimal model that supports functional and OO programming with a minimalist unified approach the \textit{Fearless Heart}.

While the notion of utilising captured state instead of fields is not entirely novel \cite{timThesis}, and there is a rich tradition of extending lambda calculus with objects \cite{finiteModelsInLC,liquori1996typed,lcObjs}, the design of  Fearless resulted in some initially counter-intuitive but ultimately powerful clarity around OO programming. At first glance, one might think that capturing state as fields is a more clear and organised approach or that both kinds of state management are ultimately equivalent. However, there is a notable difference: capturing state within a closure results in one method having access to that state, instead of every method\footnote{Fields can be used by all methods of an object and, depending on visibility, even by methods from subclasses. While this can be emulated by adding a method serving as a getter, we expect most state to be directly captured and used only by the methods actually needing it.}. This allows for a clearer and more surgical view of how data flows, helping human programmers and automated analysis tools, like type systems.

Alongside being practically usable and small, the Fearless Heart is a very extensible model. We refer to the family of extensions that can be built from the Fearless Heart as \emph{Fearless}. We explore the extensibility of the Fearless Heart by making a meaningful extension (\textit{R-Fearless}), enabling side effects and mutation while allowing equational reasoning in many situations. We do this by adapting the well-known reference capabilities/object capabilities (RC/OC) approach. It is well known that RC/OC languages can support unobservable parallelism, enforce representation invariants, and correct caching \cite{gordonRefCap,42Invariants,servetto2013balloon}, which Fearless naturally can support if it is extended with RC/OC.
We expect that other kinds of type systems tracking resources could benefit from closure-only state. R-Fearless is a non-trivial practical extension, which is still minimal enough to serve as a baseline for further RC/OC research, for example, adding more capabilities, which we discuss in our supplementary material.

This article offers the following contributions:
\begin{itemize}
    \item The Fearless Heart: a new fieldless object calculus that is practically usable for modelling programs, developing new languages and models, and writing complete real-world programs. We propose the Fearless Heart as a new baseline for extension for building object-oriented languages due to its standalone usability and logical simplicity.
    \item R-Fearless: a minimal extension supporting safe mutability with RC/OC while still being extensible and practically usable.
%\end{itemize}

%And the following minor contributions:
%Sad to comment the minor statement, but saves SO MUCH space
%\begin{itemize}
    \item A mechanised PLT-Redex \cite{redex} implementation of the Fearless Heart type system.
    \item A prototype compiler for R-Fearless with some extensions, a standard library, and an extensive suite of over 200 test programs, and more then 1800 unit/integration tests.
\end{itemize}

\textbf{Structure of the paper:}
Section~\ref{sec:core} use examples to demonstrate
the power and usability of our core model.
Section~\ref{sec:formalism} presents the formal model of the Fearless Heart.
Section~\ref{sec:braving} presents R-Fearless, providing both examples and formal definitions.
Section~\ref{sec:reasoning} discuss the advantages of R-Fearless.
The article ends with related work~\ref{sec:related} and our conclusions~\ref{sec:conclusion}.

\section{The Fearless Heart}
\label{sec:core}

\begin{wrapfigure}{r}{0.45\textwidth}
    \vspace{-8mm}
    \begin{Grammar}
        \Production{\L}{\D\OS\Xs\CS\QQ{:} \D_1\OS\Ts_1\CS, \ldots, \D_n\OS\Ts_n\CS \{\NT{\sName}\ \Ms\,\}}{}\\
        \Production{\M}{\NT{sig}\com \mid \NT{sig}\point\e\com}{}\\
        \Production{\e}{
            \x
            \mid \e\,\m\OS\Ts\CS\OR\es\CR
            \mid \L
        }{}\\
        \Production{sig}{\m\OS\Xs\CS\OR\x_1\QQ{:}\T_1\com \ldots\com \x_n\QQ{:}\T_n\CR\col\T}{}\\
    \Production{\T}{\D\OS\Ts\CS \mid \X}{}
    \end{Grammar}
    \vspace{-1mm}
    \caption{The grammar of Fearless's functional core}
    \vspace{-2mm}
    \label{fig:grammar}
\end{wrapfigure}

Fearless's functional core (Fearless Heart) is pure from the perspective of functional purity and OO purity. Those two forms of purity are conceptually very different: functional purity emphasises deterministic functions, no side effects, and immutable data, ensuring that functions consistently produce the same output for the same input and do not alter external states. On the other hand, OO purity insists that all entities in the language (including data types like numbers and booleans) are objects, providing a consistent, unified approach to data interaction through method calls.

In Figure \ref{fig:grammar}, we show the basic syntax of Fearless. Some syntactic sugar and type inference make the user-level syntax more palatable, as explained below.

% \begin{figure}[h]
%     \centering
    
% \end{figure}
A Fearless program is a sequence of top-level trait declarations \L. We use \Xs, \Ts, \Ms to represent sequences of \X, \T, and \M, respectively.
\D, \NT{X}, \NT{m} and \NT{x} are disjoint countably infinite sets.

\begin{itemize}
    \item A literal \L introduces a named trait declaration \D, parameterised by a sequence \Xs of generic type parameters and defined by a body that specifies the implemented traits, followed by a variable name representing the current instance, akin to `\Q@this@/\Q@self@' in traditional OO languages. Finally, it contains a sequence of methods \Ms.
    
\item \NT{M}: A method, either abstract (specified only by its signature \NT{sig}) or concrete (having an associated expression body \NT{e}).    
    \item \NT{e}: Expressions resemble the lambda calculus.
They can be local variables \NT{x}, conventional method calls, or object literals/closures \L.
    As you can see, top level declarations, lambdas, closures and object literals are one and the same.
    \item \NT{sig}: Method signatures adopt a conventional format, designating a method name \NT{m}, generic types \Xs, formal parameters with their types, and the return type.
    \item \NT{T}: Types can be generic type variables X or trait names instantiated with types \D\Q{[}\Ts\!\Q{]}.
\end{itemize}
To simplify scoping, no shadowing is allowed.
As a general syntactic sugar, we can omit empty parenthesis, like \lstinline@{}@ \lstinline{[]} or \lstinline{()}.
%Within expressions, \L expressions (as opposed to \L declarations), 
%\D\Q{[}\Xs\!\Q{]:}\Ts is inferred with a fresh name \D.

Overall, Fearless can be seen as a nominally typed lambda calculus, where lambdas have multiple components, and we have a table of top-level declarations labelling these lambdas.

\subsection{Introductory examples}
As the first introductory example, we could declare a Person trait with an age and a name as follows:
\begin{lstlisting}
Person:{ .age: Int, .name: Str }
\end{lstlisting}
Note how that is not a record with two fields but a trait with two methods taking zero arguments. They do not behave exactly like fields: calling those methods will trigger a (potentially non-terminating) computation. Moreover, there is no guarantee that those methods use any storage space. For example, the name can capture a string, but the age method could evaluate to the length of the same name string.
We could create an object literal that implements \Q{Person} like so:
\begin{lstlisting}
Bob:Person{.age -> 42, .name -> "Bob"}
\end{lstlisting}
However, a factory could be more appropriate, so we could write:
\begin{lstlisting}
FPerson:{
  .of(age: Num, name: Str): Person -> {.age -> age, .name -> name}
}
\end{lstlisting}
Note that we can omit `\Q@:Num@', `\Q@:Str@' and `\Q@Fresh:Person@' in \lstinline@{.age -> age, .name -> name}@. This inference is similar to what we expect from lambdas in Java, where the target type~\cite{10.1007/978-3-662-46823-4_20} is used to infer the lambda type as well as the type signature of the implemented method.
Fearless extends this inference to object literals implementing multiple methods.
Fearless also supports the inference of generic parameters for generic method calls, as with most mainstream programming languages.
Instantiation through a factory is more compact:
\Q@FPerson.of(42, "Bob")@

Note the subtle syntactic sugar in \lstinline{FPerson}, expanding to \lstinline@Fresh1[]:FPerson[]{}@. As we can see, traits with no abstract methods can be instantiated by merely referencing the trait declaration name. Indeed, this is how \lstinline{42} and \lstinline{"Bob"} work: they are just \lstinline@Fresh2[]:42[]{}@ and \lstinline@Fresh3[]:"Bob"[]{}@. We can assume a set of top-level traits implementing \lstinline{Num} and \lstinline{Str}, encoding all the numbers and strings present as literals in the code.
In this way, Fearless is a pure OO language where everything is an object, and there are no primitive types.\footnote{Of course, an implementation can optimise them away and use a more efficient implementation instead.}

\subsection{Functions and Factories}
Trait declaration names supports overloading by the arity of their generic parameters. This arity-overloading is not ambiguous since the exact number of generic arguments is known each time a declaration name is mentioned.
For the same reasons, Fearless supports method overloading based on method parameter arity.

This does not complicate the formal model; a trivial precompilation step could, for example, consistently replace all occurrences of \Q@F[A,B]@ with \Q@F/2[A,B]@, and similarly for method names.

Thus, we define a trait \lstinline{F}, which stands for both `function' and `factory'. \begin{lstlisting}
F[R]:{ #: R }
F[A, R]:{ #(a: A): R }
F[A, B, R]:{ #(a: A, b: B): R }
F[A, B, C, R]:{ #(a: A, b: B, c: C): R }
\end{lstlisting}
In the concrete syntax of Fearless, we use PascalCase for \D and \NT{X}, and we have two kinds of method names: dot-prefixed camelCase identifiers, and sequences of operator symbols like \lstinline{*} or \lstinline{<=}. 
The former is used for typical method names, while the latter suggests operators. In particular, we conventionally use the method name \lstinline{#} for the main operation of objects. We can use \lstinline{F} to rewrite factories like \lstinline{FPerson} as follows:
\begin{lstlisting}
FPerson:F[Num, Str, Person]{ age, name -> {.age -> age, .name -> name} }
\end{lstlisting}
Invocation of \lstinline{FPerson} is then even more compact: \lstinline{FPerson#(42, "Bob")}.

As we can see, when defining \lstinline{FPerson}, we can avoid mentioning the name (\lstinline{#}) of the implemented method since it is the only abstract method in \lstinline{F}. Again, this is similar to how Java does lambda inference. The concrete \L syntax requires many type annotations, but inference removes the need for most of them, resulting in compact code. Our flexible \L syntax allows us to smoothly navigate between standard lambda syntax (\lstinline@{a, b -> b}@), fully-fledged object literals, and top-level declarations.

As a variation, instead of defining \Q@Person@ and \Q@FPerson@ separately, we could define \Q{Person} inside of the method \Q{FPerson#}. This shows that names declared inside method bodies are not scoped or hidden, but are normal top level names visible everywhere. However, they can not be implemented. That is, the code below shows a compact way to models a final record-like data type.

\begin{lstlisting}
FPerson:F[Num, Str, Person]{age, name -> Person:{
  .age:Num  -> age,
  .name:Str -> name
}}
\end{lstlisting}

\subsection{Implementing booleans, optionals, and lists}

% To explore the Fearless Heart with some more involved examples, we will show the encoding of booleans, optionals, and lists in Fearless.\footnote{The encoding of Peano numbers can be found in the appendix.}
\subsubsection{Booleans}
Understanding how booleans are implemented in Fearless has proven helpful as a resource for us while explaining and teaching Fearless.

\noindent
\begin{minipage}[t]{0.48\textwidth} \begin{lstlisting}
Bool:{
  .and(other: Bool): Bool,
  .or(other: Bool): Bool,
  .not: Bool,
  .if[R](m: ThenElse[R]): R
}
ThenElse[R]:{
  .then: R,
  .else: R
}
\end{lstlisting}
\end{minipage}
\hfill
\vrule
\hfill
\begin{minipage}[t]{0.48\textwidth} \begin{lstlisting}[firstnumber=11]
True:Bool{
  .and(other) -> other,
  .or(other) -> this,
  .not -> False,
  .if(m) -> m.then,
}
False:Bool{
  .and(other) -> this,
  .or(other) -> other,
  .not -> True,
  .if(m) -> m.else,
}
\end{lstlisting}
\end{minipage}
\\[1ex]
Top-level declarations implicitly name the self-reference as \lstinline{this}. Conversely, object literals within expressions default the self-name to a fresh (unused) variable, but programmers can specify it manually if needed; indeed in many core OO calculi the self-reference name is explicit \cite{10.1007/978-3-540-45070-2_10}.

Here is an example using the booleans defined earlier:
\begin{lstlisting}
True.and(False).if({.then->"Yay", .else->"Boo",})
\end{lstlisting}
Fearless code frequently uses this pattern for representing sum types: we define a matcher trait (\lstinline{ThenElse[R]}) and a matching method (\lstinline{.if} in the example) taking an object literal implementing the matcher. % For generic containers, we generally use \lstinline{.match}, but one advantage of matching being a library feature instead of a language feature means that we can use more meaningful names like \lstinline{.if} where appropriate.
%\footnote{Keen-eyed readers may have identified that this is an implementation of the \textit{Visitor Pattern}\cite{gangOfFour}, as discussed in \autoref{sec:visitorPattern}} 
%We will discuss how Fearless enables very compact usage of the visitor pattern more in \autoref{sec:visitorPattern}.
Fearless allows the omission of round parentheses for single-argument methods,
allowing the former code to look as follows: %This can be combined with standard and intuitive indentation to simplify the code and make it easy to follow:

\begin{lstlisting}
True .and False .if {
  .then -> "Yay",
  .else -> "Boo",
}
\end{lstlisting}
By allowing to omit those parenthesis, code like `\lstinline{1 + 2 * 3}' is interpreted as a set of method calls. To avoid precedence ambiguity, we interpret all method calls as left-associative, rendering the code equivalent to` \lstinline{(1+2)*3}' rather than `\lstinline{1+(2*3)}'.
This intentional design choice, while unconventional, greatly simplifies the use of fluent interfaces \cite{fluentInterface}, as we will demonstrate later. 
As usual, we can always use parentheses to control precedence.

\subsubsection{Optionals}
\label{sec:optionals}
Optionals are conceptually similar to booleans, with the distinction being that they can capture a value in their `\Q@.some@' case. The matcher is generic on the type of this value:
\begin{lstlisting}
Opt[T]:{
  .match[R](m: OptMatch[T,R]): R -> m.empty 
}
OptMatch[T,R]:{
  .empty: R,
  .some(t: T): R
}
Opt:{ #[T](t: T): Opt[T] -> {m -> m.some(t)} }
\end{lstlisting}
Then, we can instantiate an empty optional containing our bob person from earlier with \lstinline{Opt#bob}. An empty optional could be instantiated like \lstinline{Opt[Person]}, or \lstinline@{}@ thanks to inference if the empty optional is passed as a method parameter or is used as a return value.

\subsubsection{Lists}
A linked list can be seen as optionals where the `\Q@.some@/\Q@.elem@' case holds two pieces of information: the remainder of the list and a current final element.
\begin{lstlisting}
List[T]:{
  .match[R](m: ListMatch[T,R]): R -> m.empty
  +(e: T): List[T] -> {m -> m.elem(this, e)},
}
ListMatch[T,R]:{
  .empty: R,
  .elem(list: List[T], e: T): R
}
\end{lstlisting}
We can produce the list \lstinline{[1,2,3]} by writing \lstinline{List[Num]+1+2+3}, where \lstinline{List[Num]} is the empty list.
More involved kinds of lists are also expressible as shown in the code below. Note how \Q@{}@ is used for both empty optionals and empty lists thanks to target type inference \cite{10.1007/978-3-662-46823-4_20}.
\begin{lstlisting}
List[Opt[Num]]+{}+{}+(Opt#3) // [empty, empty, some(3)]
List[List[Num]]+{}+{}+(List[Num]+3) // [[], [], [3]]
\end{lstlisting}
With these encodings, we can illustrate some practical examples. For instance, we could compute the sum of all the elements in a list:
\begin{lstlisting}
Example:{ .sum(ns: List[Num]): Num -> ns.match{
    .empty -> 0,
    .elem(list, e) -> this.sum(list) + e
}}
\end{lstlisting}
We can also add a \lstinline{.map} method to \lstinline{List}:
\begin{lstlisting}
List[T]:{..//as before
  .map[R](f: F[T,R]): List[R] -> this.match{
    .empty -> {},
    .elem(list, e) -> list.map(f) + (f#e)
}}
\end{lstlisting}
% While there are a few extra convenience methods in the Fearless standard library, the examples shown here are largely identical to those in our standard library.

\subsubsection{Data Representation}
% TODO: Should this whole subsection be moved or killed?
Traditional OO design portrays entities like \mbox{\lstinline{List},} \mbox{\lstinline{EmptyList},} and \lstinline{NonEmptyList} as separate classes, each explicitly presenting their respective fields and behaviours. In contrast, Fearless encourages consolidating different list types into a single List trait, differentiating the cases through methods within the \lstinline{ListMatch[T,R]} trait:
\lstinline{.elem(list: List[T], e: T)} and
\lstinline{.empty}. Here, we express data variants and the state maintained in each variant through method signatures rather than separate top-level declarations.
The same discussion holds for \Q@Opt@ as well.
Moving this information from the top-level definitions into method signatures represents a different mindset, 
radically different from the classical OO design philosophy, embodied in a graphical representation by UML class diagrams.
%We leave exploring those considerations to future work; in this article, we focus on the language itself.

\subsection{The Visitor Pattern as a match}
\label{sec:visitorPattern}
In OO languages, the
\textit{Visitor Pattern} \cite{gangOfFour}
separates behaviour from data variants. By making operations into objects, we can dynamically add operations without modifying the data variants.
Let us try to design a more complex data structure as a design exercise to explore the features of Fearless that we just covered. Consider representing a subset of HTML as an AST. For simplicity, we chose to represent only headers (h1, h5), anchor tags (a), and divisions (div). We start by defining a trait, \lstinline{Html}, and its matcher trait, \lstinline{HtmlMatch[R]}.
We then define a convenient factory class \lstinline{FHtml} with a method for each data variant, linking it to the method with the same name in the matcher.

\noindent
\begin{minipage}[t]{0.485\textwidth} \begin{lstlisting}
Html:{.match[R](m: HtmlMatch[R]): R}
HtmlMatch[R]:{
  .h1(text: Str): R,
  .h5(text: Str): R,
  .a(link: Str, text: Str): R,
  .div(es: List[Html]): R,
}
\end{lstlisting}
\end{minipage}\!\!
\hfill
\vrule
\hfill
\begin{minipage}[t]{0.50\textwidth} \begin{lstlisting}[firstnumber=8]
FHtml:HtmlMatch[Html]{
  .h1(text) -> {m -> m.h1 text},
  .h5(text) -> {m -> m.h5 text},
  .a(l, text) -> {m -> m.a(l, text)},
  .div(es) -> {m -> m.div es},
}
\end{lstlisting}
\end{minipage}
\\[1ex]

These methods, such as \lstinline{.h1(text: Str): R} and \lstinline{.a(link: Str, text: Str): R}, deconstruct the various HTML elements providing the data they encapsulate.

We were surprised when we discovered that the \lstinline{FHtml} factory can be defined by implementing \lstinline{HtmlMatch[Html]}.
Indeed, \Q{FHtml} is a shallow-clone visitor, a trait which encapsulates the logic to create a shallow clone of an HTML entity.
We can easily define a deep-clone visitor that explores all the elements by extending the shallow-clone visitor/factory and overriding all the composite cases \cite{gangOfFour}.
For our simple example, the only composite is the div:
\begin{lstlisting}
HtmlCloneVisitor:FHtml{ .div(es) -> FHtml.div(es.map{ e -> e.match this }) }
\end{lstlisting}

At first sight, the clone visitor is a pointless operation since it creates a structurally identical copy of a deeply immutable data type. However, every visitor method works as a handler for the template method pattern. This means that an object literal can implement the clone visitor trait and only define cases for the relevant parts of the AST. Indeed, visitors shine because they can use inheritance/code reuse to produce complex behaviour in a very compact way. 

Here, we can see how our type inference makes the clone visitor much more compact than the equivalent version in, for example, Java. This is counterintuitive since, from a technical perspective, our inference is \textbf{not} more sophisticated or advanced than that of Java.

The \lstinline{CapitaliseTitles} trait, extending \lstinline{HtmlCloneVisitor}, relies on the handler \lstinline{.h1} to override the behaviour and transform h1 headers by turning their text to uppercase:
\begin{lstlisting}
CapitaliseTitles:HtmlCloneVisitor{ .h1(text) -> FHtml.h1(text.upperCase) }
 ... myHtml.match(CapitaliseTitles) ... //usage
// or, a direct definition without needing the CapitaliseTitles trait:
myHtml.match{.h1(text) -> FHtml.h1(text.upperCase)}
\end{lstlisting}
% In this example, where the \lstinline{CapitalizeTitles} reuses the logic of all the cases except for \lstinline{.h1}, it becomes clear that Fearless traits enable code reuse similarly to how interfaces with default implementations and multiple inheritance do in languages like Java 8 or C\#.

\noindent In the future we could examine the correlation between \lstinline{FHtml} as a visitor and object algebras \cite{objectAlgebras}.

\subsection{Why real-world usability matters}
Consider someone building a new language by extending FJ and adding some dataflow analysis for secure programming \cite{toplas42}. The hand-wave introduction of features like local variables, if-statements or while-loops would introduce new forms of information leakage unaddressed by the original model.
Those other features are usually only hand-waved in or introduced in different formalisations and informally merged. 
This may be brittle and could be a source of some hidden unsoundness, as it was the case with the combination of \Q[language=Java]@null@ and wildcard bounds \cite{10.1145/2983990.2984004}.
The Fearless Heart avoids this problem by making it trivial to encode user-friendly support for all of these features at the library level with relatively small amounts of code. Moving these language features to the library level means the Fearless Heart remains simple. Additionally, implementing a feature like booleans with if-else in our model is not just writing a program; it is a formal model of that language feature by virtue of it being written in the syntax and semantics of the Fearless Heart's formal model.

%I removed this becase -1 normally PLs do not have visitors in the language but is normal to have them in libraries, -2 it is one line only beacuse in our artificial example we have one one composite node
%The previous section's demonstration of our 1-line deep clone visitor was not just a demonstration of how the Fearless Heart lets us use design patterns with minimal syntax. It was also effectively a 1-line formal model of a deep clone visitor.

\section{Formalising the Fearless Heart}
\label{sec:formalism}
\input{formalism}

\section{Braving Mutability: Fearless's Journey into Mutability}
\label{sec:braving}
The Fearless Heart is an expression-based programming language. However, with the introduction of mutability, some expressions are executed to cause a side effect and have no meaningful return value. For these expressions, we can add \lstinline{Void} and \lstinline{Block} traits. Fearless allows \QQ{\_} as an unused variable name, and is desugared to a fresh name.
\begin{lstlisting}
Void:{}
Block:{
  #[A,B](_: A, res: B): B -> res,
  #[A,B,C](_: A, _: B, res: C): C -> res, //and a few more overloads
}
\end{lstlisting}
To enable mutability, we extend Fearless with type \lstinline{Ref[T]} and its magic factory type. It is magic because the user cannot express the expected functionality of \lstinline{Ref[T]} in the language. Our magic is similar to how OCaml handles mutation. In a first approximation, their signatures look like this:
\begin{lstlisting}
Ref[T]: {.get: T, .set(x: T): Void }//First approximation, not ok yet 
Ref: { #[T](x: T): Ref[T] -> Magic! }
\end{lstlisting}
The idea is that instances of \lstinline{Ref[T]} produced with the \lstinline{Ref#} method would behave as objects with a single non-final field of type T, that can be read or updated.
However, this can easily be misused to create rampant aliasing and pervasive mutations, preventing equational reasoning.
We avoid those issues using  RC/OC, as introduced by \cite{ponyCaps,42Recovery,gordonRefCap}.
In particular, Pony \cite{ponyCaps} is the most mature approach using RC/OC.
Our reference capability system aims to retain all the important expressiveness from Pony, while being overall substantially simpler. Pony has six core capabilities which can be combined in different ways giving 72 different possible combinations \cite{nobleRefCapsRegions}.
% R-Fearless has four reference capabilities, and the interaction between those capabilities are simpler.

\subsection{Reference capabilities in a nutshell}
The main idea of reference capabilities is to divide objects into two kinds: mutable and immutable objects.
The reachable object graph (ROG) of an \textbf{immutable} object can not be mutated. All objects in the ROG of an immutable object are immutable (deep immutability with no interior mutability backdoors). These rules do not prevent \lstinline{Ref[T]} from appearing in the ROG of immutable objects, but they require that the .set method is uncallable on those \lstinline{Ref[T]}; they are frozen.

A \textbf{mutable} object can have both mutable and immutable objects in its ROG. We use the term `mutable ROG' (MROG) of an object to indicate the mutable objects reachable from a certain object. Mutable objects can become immutable, while immutable objects are permanently immutable.

Crucially, reference capability type systems do not directly track mutable and immutable objects, but track the references to such objects by introducing (in our case) four reference capabilities: \Q{imm} (immutable), \Q{iso} (isolated), \Q{read} (readable), and \Q{mut} (mutable).

%\begin{Grammar}
%    \Production{R}{\QQ{imm} %\mid \QQ{iso} \mid \QQ{read} %\mid \QQ{mut}}{}
%\end{Grammar}

\lstinline{iso} and \lstinline{imm} are pure; they allow equational reasoning: 
\begin{itemize}
    \item An \lstinline{iso} refers to a mutable object whose MROG is only reachable from that reference. 
    \item An \lstinline{imm} refers to a deeply immutable object.

\end{itemize}

\lstinline{iso} references are affine: they can only be used zero or one time in their scope. The main feature of \lstinline{iso} references is that they are a subtype of \lstinline{mut} and \lstinline{imm} references. In this way, they can be used to create immutable objects, or they can be opened up into conventional mutable references.
The main feature of \Q{imm} references is that they are not part of MROGs, thus they can be freely aliased between isolated portions of the object graph.
This is a more implicit version of what happens in \cite{nobleRefCapsRegions}, where immutable objects are explicitly placed outside of all of the regions.

\lstinline{read} and \lstinline{mut} are impure; extra care must be taken when reasoning about their referred values:
\begin{itemize}
    \item A \lstinline{mut}  reference points to a mutable object. The MROG of that object can be mutated using this reference in accordance with the API of its type. Mutable references are akin to unrestricted references in languages like C\#, Java, and Swift. Our \Q{mut} is unlike \Q{&mut} in Rust because we allow multiple mutable aliases to the same object.
% \footnote{\Q@mut@ owned values in Rust are more similar to \Q{iso} references than to mut references.} % I don't think we need to say this.

    \item A \lstinline{read} reference points to either a mutable or an immutable object; thus, the MROG of that object can not be mutated using this reference, but it could be mutated through some other alias. Readable refences are the common super-type of all kinds of references.
\end{itemize}

\subsection{R-Fearless}
\begin{wrapfigure}{r}{0.45\textwidth}
    \vspace{-8mm}
    \begin{Grammar}
        \Production{L}{\D\OS\Xs\QQ{]:} \NT{D_1}\OS\Ts_1\CS, \ldots, \D_n\OS\Ts_n\CS \{\NT{\sName}\ \Ms\}}{}\\
        \Production{M}{\NT{sig\com} \mid \NT{sig}\quad \point\ \NT{e\com}}{}\\
        \Production{e}{
            \NT{x}
            \mid \NT{e}\fstop\NT{m}\OS\Ts\CS\OR\NT{\overline{e}}\QQ{)}
            \mid \NT{R}\ \L
        }{}\\
        \Production{sig}{\NT{R}\ \NT{m}\OS\Xs\CS\OR\NT{x_1\QQ{:}T_1, \ldots, x_n\QQ{:}T_n}\CR\col\NT{T}}{}\\
    \Production{T}{
        \NT{R}\ \NT{D\OS\overline{T}\CS}
        \mid \NT{X} \mid
        \NT{R}\ \NT{X}}{}\\
    \Production{R}{\QQ{imm} \mid \QQ{iso} \mid \QQ{read} \mid \QQ{mut}}{}
    \end{Grammar}
    \vspace{-1mm}
    \caption{The grammar of Fearless with reference capabilities}
    \vspace{-2mm}
    \label{fig:capsGrammar}
\end{wrapfigure}
We now define R-Fearless, an extension of Fearless Heart with RC/OC, and we show the interesting parts of how to extend the formalisation.
For space reasons, we provide a full formalisation of R-Fearless only in our supplementary material.
In Fig. \ref{fig:capsGrammar}, we can see the language grammar, including reference capabilities. Notably, we start the method signature \NT{sig} with the reference capability of the receiver, and we start the object literal with the capability of its initial reference.

We now have three kinds of types: concrete types with their reference capabilities ($\R\,\NT{D\OS\Ts\CS}$), type variables \NT{X}  inheriting the capability of the type instantiating them, and type variables with an explicit capability ($\NT{R}\ \NT{X}$).

Handling genericity with reference capabilities opens some design questions. Should the genericity be about the trait name, the capabilities, the types or all of the above?
In our chosen design, a type parameter \NT{X} will be instantiated with the pair
$\NT{R}\ \NT{D\OS\Ts\CS}$. To override the first element of the pair we use the syntax $\NT{R}\ \NT{X}$ (i.e. \lstinline{imm X}) to provide a specific reference capability \NT{R} with the provided \NT{D\OS\Ts\CS} instead of using the one attached initially to that type. This design decision allows for code that is parametric over type names and capabilities and for code that is solely parametric over type names.
R-Fearless needs to add two more well formedness criteria, related to the \iso being affine:
First, a parameter of type $\iso\,\_$ can be used only one time in the method body, or any number of times but only inside object literals. Second, in a valid $\D\OS\Ts\CS$ any $\T\in\Ts$ is not of form $\iso\, \_$.

In practical programs, most references are \lstinline{imm}, so in the concrete fearless syntax, we consider that every time a reference capability is omitted on a \NT{D\OS\Ts\CS} or on a method signature, \lstinline{imm} is inferred. In this way, all the code shown in \autoref{sec:core} is still valid Fearless code. For example, consider the code below:
\begin{lstlisting}
F[A,R]:{ #(a: A): R, } //shown earlier, imm receiver, handles any kind of A,R
Ex[A]:{mut #(a: imm A, b: mut A): Person,}//mut rec, imm A*mut A->imm Person
\end{lstlisting}
% In the \lstinline{F} shown earlier, \lstinline{A} and \lstinline{R} can refer to types with any reference capability for both \lstinline{A} and \lstinline{R}. On the other hand, in \lstinline{MutToImm}, the reference capability provided as part of the type argument for \lstinline{A} is replaced with \lstinline{mut}, requiring `\lstinline{a}' to be \lstinline{mut}, and guaranteeing that the return value will always be \Q{imm}.

\subsection{\lstinline{Ref[T]} with reference capabilities}
 With our reference capabilities, \lstinline{Ref[T]} can be written as follows:
\begin{lstlisting}
Ref[T]:{
  mut  .get: T,
  read .rget: read T,
  mut  .swap(x: T): T,
  mut  .set(x: T): Void -> Block#(this.swap(x), Void),
}
Ref:{ #[T](x: T): mut Ref[T] -> Magic! }
\end{lstlisting}
% TODO: I skipped the updated definition of Void and Block because I think it's obvious and redundant. I've also skipped discussion of Sealed.

Crucially, we now have two different getters. On a \lstinline{mut} receiver, we return the value as indicated by the generic type parameter. In contrast, on a \lstinline{read} receiver, we return the value as \lstinline{read}.

Another Fearless extension could allow overloading on the receiver reference, allowing both methods to be called `\Q{.get}'.
That overloading extension would cause the semantics to depend on the type system. 
%To keep R-Fearless extensible for use cases that do not desire that level of overloading, we have not applied this extension to R-Fearless.
%No, this is redundant and too defensive
We leave to future work a discussion on the pros and cons of this decision, the details of our overloading resolution, and the impact on type inference. Our prototype compiler, which implements an extension of R-Fearless, currently supports this overloading.
%This approach allows us to avoid needing "viewpoint adapted types"\cite{arrowTypesPony} when writing code parametric on multiple modifiers and viewpoint adaptations.

\subsection{Promotions}
\begin{wrapfigure}{r}{0.3\textwidth}
    \centering
    \vspace{-8mm}\includegraphics[width=0.9\linewidth]{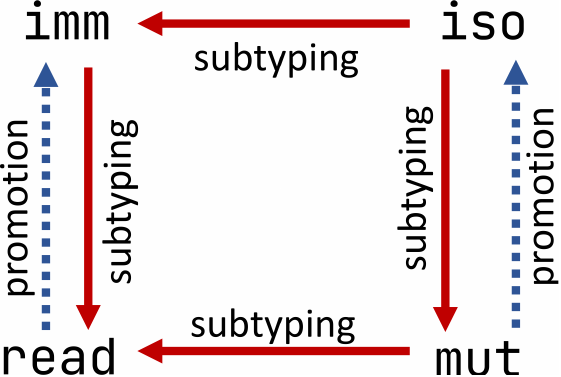}
    \caption{Relationships between capabilities in R-Fearless.}
    \Description{A diagram showing that read is a sub-type of imm, and that read is promotable to imm. The diagram also shows that imm and mut are a sub-types of iso, and that mut is promotable to iso. Finally, the diagram shows that read is a subtype of mut.}
    \vspace{-1mm}
    \label{fig:caps}    \vspace{-2mm}
\end{wrapfigure}
While the initial reference to an object could be of any kind, reference capabilities encourage a programming style where objects are born mutable, go through some mutation, and then become immutable via a `promotion/recovery' mechanism.

Reference capabilities are very flexible and can be described as `opt-in': 
A programmer could chose to use \lstinline{imm} for strings, booleans, numbers and other immutable data types, and \lstinline{mut} in any other case. Then another programmer would be able to take their work and \emph{promote/recover} the mutable objects from the first programmer's code to \lstinline{iso} or to \lstinline{imm}, even with the first programmer not doing any work to make this possible.

We can illustrate how promotion/recovery works with a few code examples using \lstinline{Ref}. Note that the \lstinline{Ref[T]} type is not treated specially by the type system, and it is not magic (only the \lstinline{Ref} factory is). Our examples would work identically with any other container-like trait.
\begin{lstlisting}
.ex01(r: Ref[Person], p: Person): Void -> r.set(p), //error: r is imm
.ex02(r: read Ref[Person], p: Person): Void -> r.set(p), //error: r is read
.ex03(r: mut Ref[Person], p: Person):Void -> r.set(p) //ok
.ex04(r: mut Ref[Person]): Ref[Person] -> r //error: subtyping
.ex05(r: mut Ref[Person]): read Ref[Person] -> r //ok, subtyping
\end{lstlisting}
The first three examples show that mutable methods can only be called on mutable (and isolated) receivers. \lstinline{.ex04} and \lstinline{.ex05} show allowed sub-typing.
\begin{lstlisting}
.ex06(p: Person): mut Ref[Person] -> Ref#p //ok
.ex07(p: Person): iso Ref[Person] -> Ref#p //ok, mut promoted to iso
.ex08(p: Person):     Ref[Person] -> Ref#p //ok
\end{lstlisting}
Those three examples show promotions of object creations: \lstinline{.ex06} shows the type directly returned by the factory, \lstinline{.ex07} shows how \mut can be promoted to \iso: since no actual parameters passed to the \lstinline{#} method call are \mut/\read, the \mut result must have been created internally; thus, we can promote it to \iso.
\lstinline{.ex08} shows that \iso promotion + subtyping is a common way to create immutable objects from mutable ones. 
\begin{lstlisting}
.ex09(r: mut Ref[Person]): Person -> r.get, //ok
.ex10(r: Ref[Person]): read Person-> r.rget, //ok
.ex11(r: Ref[Person]): Person -> r.rget, //ok, read promotion to imm
.ex12(r: read Ref[Person]): Person -> r.rget //error :(
\end{lstlisting}
Those four examples show calls to getters.
\lstinline{.ex09} shows a direct call to \lstinline{.get} obtaining the result as declared in \lstinline{.get}.
\lstinline{.ex10} shows a direct call to \lstinline{.rget} obtaining the type as declared in \lstinline{.rget}. However, it is undesirable to get the result as \lstinline{read} because we statically know that the stored person is \lstinline{imm}, which is a more useful capability.
\lstinline{.ex11} shows that we can indeed return an \lstinline{imm Person}:
because \lstinline{r} is \lstinline{imm}, no actual parameters for the call \lstinline{r.rget} are \lstinline{mut}/\lstinline{read}, thus the \read result must either be \imm, or be an internally created \mut object that can be promoted to \imm as for \lstinline{.ex08}.
Promotion relies on the type of the actual parameters of the call site and not on the formal parameters from the method declaration.

As shown in the prior examples, defining two kinds of getters (one for \lstinline{read} and one for \lstinline{mut}) covers many use cases, including all the cases for non-generic containers.
However, \lstinline{Ref} is a generic container, and the interaction of reference capabilities with generics is complex. This is shown in the \lstinline{.ex12} example: we have a \lstinline{read} reference to a \lstinline{Ref} containing an \lstinline{imm Person}. We can not call the \lstinline{.get} method that would return an \lstinline{imm} because that requires a \lstinline{mut} \lstinline{Ref}. We also cannot apply promotion because the method capability for \lstinline{.rget} is \read. One may think that we statically know that our \lstinline{read Ref} of \lstinline{imm Person} must contain a single \lstinline{imm Person}, but this is not the case, consider the following example:
\begin{lstlisting}
.break(bob: imm Person, adam: mut Person): read Ref[imm Person] -> {'self
  .get -> bob,   //bob is imm
  .rget -> adam, // adam is not imm! allowed: rget returns read
  .swap(x) -> self.swap(x), // loop
}
\end{lstlisting}
In an alternative version of R-Fearless allowing \lstinline{.ex12}, if \lstinline{.ex12} was called with the result of \lstinline{.break}, it would unsoundly produce an \lstinline{imm Person} reference pointing to a \Q{mut Person}. We are considering various options to allow for a more expressive signature for \lstinline{.rget} to capture the user intent better. Pony solves this with viewpoint types \cite{arrowTypesPony} and M\# solves this with deferred permission composition \cite{gordonRefCap}; which R-Fearless could be trivially extended to support, but we are currently trying to find a more minimal solution.

This limitation can currently be circumvented in R-Fearless by using an extra level of indirection:
\begin{lstlisting}
BoxI[T]:{ read #: imm T }
Examples:{ .ex12Fixed(r: read Ref[BoxI[Person]]): Person -> r.rget#, }
\end{lstlisting}

\subsubsection{Formally modelling promotion}
Promotion works as if any method had an additional valid signature, where all the \lstinline{mut}s are turned into \lstinline{iso}s and all the \lstinline{read}s are turned into \lstinline{imm}s.
On the parameters side, this makes the method harder to call. On the return type side, this makes the method result more useful. 
%This is also implicitly applied to whole method bodies: if a method body does not use any \lstinline{mut}/\lstinline{read}/transparent generic parameters, then it can return a \lstinline{mut} value as an \lstinline{iso} or \lstinline{imm}, and a \lstinline{read} value as an \lstinline{imm}.
We can model this formally by adding another typing rule  
\textsc{prom-call-T}, which supports our promoted calls.

\metarule{prom-call-t}{
    \Xs', \Gamma \vdash \e_0 : \R_0\, \D\OS\Ts_0\CS\qquad \R_0 \in \{\imm,\iso\} \qquad \R_0 \leq \R
    \qquad \Xs'\text{disj}\, \Xs\\
    
    \R\ \m\OS\Xs\CS\OR\x_1\QQ{:}\T_1, \ldots, \x_n\QQ{:}\T_n\CR\col\T\ \ \point \e\com \in \mathrm{meths}(\R_0\,\D\OS\Ts_0\CS)\\
    \Xs', \Gamma \vdash \e_1 : \PromP{\T_1[\Xs = \Ts]}\ \ldots\ \Xs', \Gamma \vdash \e_n : \PromP{\T_n[\Xs = \Ts]}
}{
    \Xs', \Gamma \vdash \e_0\,\m\OS\Ts\CS\OR\e_1,\ldots\e_n\CR : 
    \PromR{\T[\Xs = \Ts]}
}{}\\[1ex]

This rule is only applicable if the receiver is \imm or \iso, and uses the notations defined below to apply the promotion. \mut and \read are promoted into \iso and \imm, and promotion on types propagates to the \R. However, types of form \X do not have an \R component. They need to be handled defensively, since they could be instantiated with any \R. Thus, for parameters we require them to be \iso and for the return type we only allow them to be \imm.

\begin{Define}{8}{0}{
    \Prom\R = \R'\qquad
    \PromP\T = \T'\qquad
    \PromR\T = \T'\qquad
}
  \With{
    \Prom\mut = \iso\qquad
    \Prom\read = \imm\qquad
    \Prom\R = \R\ \mathrm{where}\ \R \notin \{ \mut, \read \}
  }{}

\With{
    \PromP{\R\, \D\OS\Ts\CS} = \Prom\R\ \D\OS\Ts\CS\qquad
    \PromP{\R\, \X} = \Prom\R\ \X\qquad
    \PromP\X = \iso\, X
}{}

\With{
    \PromR{\R\, \D\OS\Ts\CS} = \Prom\R\ \D\OS\Ts\CS\qquad
    \,\,\PromR{\R\, \X} = \Prom\R\ \X\qquad
    \PromR\X = \imm\,\X
}{}
\end{Define}

Reference capabilities are well known, and there are plenty of examples in the literature showing their flexibility and their soundness properties \cite{castegren2016reference,gordonRefCap,javari,10.1145/1035292.1028980,10.1145/2824815.2824816,giannini2016types}. We omit repeating those examples and proofs for space reasons.

\subsection{Capturing RC in closures}
One crucial point of R-Fearless is how we support capturing reference capabilities into closures instead of supporting fields.
The key insight is that applying RC to fields is complicated and is not just a straightforward propagation of applying the capabilities to references.
There are two non-obvious relations between fields and RC:

First, a \lstinline{mut} reference always refers to a \lstinline{mut} object; should a \lstinline{mut} field always refer to a \lstinline{mut} object? Confusingly, the answer is no. Fields are (usually) declared in classes, while instances of classes can be \lstinline{mut} and \lstinline{imm} objects. The whole ROG of an \lstinline{imm} object is \lstinline{imm}, so even the field initially declared as \lstinline{mut} in the class will hold an \lstinline{imm} object when the receiver is \lstinline{imm}.
    That is, \mut fields are more akin to \lstinline{RoMaybe}/\lstinline{Poly}/\lstinline{PolyRead} fields \cite{pluggableTypes,javari}.
    
Second, an \lstinline{iso} reference is affine. Would an \lstinline{iso} field be affine too? Affine fields are not used in RC literature. Instead, \lstinline{iso} fields (if allowed at all) have some different behaviour, often destructive reads plus complex language-supported patterns where the field is automatically consumed and then repristinated.

Those counter-intuitive factors make it difficult to explain an RC system to programmers and other researchers.
Since Fearless has no fields, a programming guide could avoid this complexity and simply move on, and a formalisation and proof would not need to model those different regimes. Moreover, closures are now widespread in modern programming languages, so a realistic RC/OC system supporting fields would still need to also support closures.

In a non-RC/OC system, an object literal could see all the variables currently in scope. This is formally captured in \textsc{lit-ok}, where
a method $M$ of the literal is typed in the full unrestricted outer $\Gamma$ environment plus the self type.
Instead, in R-Fearless, the method will see an adapted $\Gamma_i$ as shown below,
%$ = (\Gamma, x:D[Xs])[R_0,R]$, where
%$R_0$ is the RC of the object literal, and \R is the RC of the method.
where `rcOf$(\M)$' gives the receiver \R of a method \M:
%, and callable and $\Gamma[\R_0,\R_1]$:

\metarule{r-lit-ok}{
    \Gamma_i = (\Gamma, \x:\R_0\,\D\OS\Xs\CS)[\R_0,\mathrm{rcOf}(\M_i)]\qquad
    \Xs, \Gamma_1 \vdash \M_1 : \mathrm{OK}\ 
    \ \ldots\ 
    \Xs, \Gamma_n\vdash \M_n : \mathrm{OK}\\
    \mathrm{overrideOk}(\D\OS\Xs\CS)\qquad
    \mathrm{implementOk}(\D\OS\Xs\CS)
    %\mathrm{meths}(\D\OS\Xs\CS)
}{
    \Gamma \vdash \R_0\ \D\OS\Xs\QQ{]:} \_\  \{\NT{\sName}\ \NT{M}_1 \ldots \NT{M}_n \} : \mathrm{OK}
}{}

\begin{Define}{24}{0}{
    \Gamma[R_0,R] = \Gamma'\qquad \T[\R]=\T'
}
%Case1
  \Line{(\x: \T, \Gamma)[\R_0,\R_1]
  }{
  \x: \T[\imm],\ \   \Gamma[\R_0,\R_1]
  \quad \text{ with }  \T = \iso \_ \text{ or } \imm \_
  }{}
%Case2
  \Line{(\x: \T, \Gamma)[\R_0,\R_1]
  }{
  \x: \T,\ \   \Gamma[\R_0,\R_1]
  \quad \text{ with }  \R_0\in\{\iso,\mut\}
  \text{ and } \R_1\in\{\iso,\mut\}
  }{}
%Case3
  \Line{(\x: \T, \Gamma)[\R_0,\imm]
  }{
  \x: \T[\imm],\ \   \Gamma[\R_0,\imm]
  \quad \text{ with }  \R_0\in\{\iso,\mut,\read\}
  }{}
%Case4
\Line{(\x: \T, \Gamma)[\R_0,\read]
  }{
  \x: \T[\read],\ \   \Gamma[\R_0,\read]
  \quad \text{ with }  \R_0\in\{\iso,\mut,\read\}
  }{}
%CaseEmpty
\Line{\emptyset[\R_0,\R_1]
  }{\emptyset
  }{}
%T[R]
\With{
\R\, \D[\Ts][\R'] = \R'\D[\Ts]\qquad
\R\, \X[\R'] = \R'\X\qquad
\X[\R'] = \R'\X
}{}
\end{Define}
%also add back the notation for T[R]
Our notation for adapting\cite{genericUniverseTypes} $\Gamma$ is built upon the intuition that while fields are visible from all the methods, variables are captured by individual method bodies. Therefore, the method's RC offers more information about how and when a captured variable could be used, and this extra information allows for more flexible typing than RC fields.

The above notation defines a set of cases where the first one matching is to be applied.
First, \iso and \imm variables are always visible, and the method sees them as \imm.
\iso variables cannot be seen as \iso, since the method could be called multiple times. 
An alternative language design could capture them as either \imm or \mut depending on the way the method use them, but this would make the language less predictable without introducing any real benefit: if the programmer wants to capture them as \mut, they can simply use subtyping ($\iso\leq\mut$).

Then, if both the receiver and the method RCs are \iso or \mut, the variable is seen as it is in $\Gamma$.

The next case handles \imm methods.
One may assume that an \imm method can only capture \imm and \iso variables, and this intuition holds for \imm literals.
But if the literal is \iso/\mut/\read, something very interesting happens: 
since whole ROGs are promoted, instead of individual objects, the method cannot be called until the object (and its whole ROG) becomes immutable.
This means that all the captured state will not be used until the point in time where all the captured state has become immutable. Thus, from a type system perspective, we can relax the capturing constraints and see all the captured variables (even \mut ones) as \imm. The method will only be callable in a context where there are no more \mut aliases to the captured state.
Pony exhibits this same exact capturing behaviour \cite{ponyCaps}.

The next case instead restricts the type of the captured variables in the case of \read methods: even if a \read method captures a \mut  variable, it will see it as \read, in order to prevent mutation to originate from a \read receiver.

%Finally, if no other case apply (Nick, is there any situation where we discard the variable in RFearless??? I do not think so! Nick: There is no case where we filter out variables with our 4 modifiers) we simply discard the variable.

% \subsection{Playing with \lstinline{iso}}

\subsection{Callable and implementable methods with RC}
In an OO language without RC, any reference to an object of type \NT{T} can access the whole API of the type \NT{T}. The main point of RC is that they restrict the set of available methods so that from an \lstinline{imm} reference, we can call \lstinline{imm} methods, and from a \lstinline{mut} reference, we can call \lstinline{mut} methods. \lstinline{iso} references can call all kinds of methods, and \lstinline{read} methods can be called via any kind of reference.
Conceptually, an immutable object does not need any \mut methods at all. When a reference is promoted to \lstinline{imm}, any \mut methods may cease to exist because it is guaranteed that they will never be callable again.
For this reason, \imm and \read object literals can leave all \mut and \iso methods inherited from implemented traits abstract.

One of the fundamental ideas in class-based object oriented languages is that all objects of a certain class offer (at least) all the methods of that class.

Reference capabilities break this idea, and instead partition the set of methods into groups, and immutable objects only have \imm/\read methods.
Mutable objects conceptually have all kinds of methods since they can become immutable via promotion.
Formally, `$\mathrm{callable}(\R, \M)$ iff either $\R\in\{\mut,\iso\}$ or  rcOf$(\M)\in\{\imm,\read\}$'.
This consideration can be used to relax typing:
an object literal whose original reference is \imm or \read, need not to implement \mut and \iso methods.
This is implemented in the formalism of R-Fearless by changing the rule \textsc{lit-t} as shown below.

\metarule{r-lit-t}{
    \Xs \subseteq \Xs'\qquad
    \forall \NT{M} \in \mathrm{meths}(\D\OS\Xs\CS), \ \text{if callable}(\R,\M) \text{ then not abs}(\M)\\
    \L=\R\ \D\OS\Xs\CS\QQ{:} \_ \ \OC\sName\ \Ms\CC\qquad
    \forall \M \in \Ms, \ \text{callable}(\R,\M)\qquad
    \Gamma\vdash
    \L : \mathrm{OK}
}{
    \Xs',\ \Gamma \vdash \R\ \L : \R\ \D\OS\Xs\CS
}{}

This rule states that only the callable methods must be not abstract, and that uncallable methods can not be defined: they would just be dead code if allowed.
%\begin{Define}{0}{0}{
%    \mathrm{callable}(\R, \M) \text{ iff }\R\in\{\QQ{imm},\QQ{read}\} \text{ implies  rcOf}(\M)\not\in\{\QQ{mut},\QQ{iso}\}
%}
%\end{Define}
%All methods are potentially `callable' from a receiver $\R_0$ if the receiver is \mut or \iso (because via promotion/sub-typing they can become any other RC). However, if the receiver is \imm or \read, then \mut and \iso methods would be uncallable.

%From a runtime perspective, it could be possible to garbage collect all objects that are only reachable via uncallable methods.
%However, extensive experimentation would be needed to determine if such an optimisation would be beneficial in practical use cases.
% Investigating if such an optimisation is actually beneficial in practical use cases would require a significant time investment.

%This alternative signature will replace all the have version of the same $\M$

\subsection{Object capabilities}
Object Capabilities (OC) \cite{10.1145/361932.361937,objCapsE} are not a type system feature but a programming methodology that is greatly beneficial when the standard library embraces it.
The main idea is that instead of being able to perform non-deterministic actions like IO everywhere in the code by using static methods or public constructors, only specific objects have the `capability' of doing those privileged actions, and access to those objects is kept under strict control. We do this in R-Fearless with a \lstinline{mut System} object.

\lstinline{System} is a normal trait with no implemented methods. An instance of \lstinline{System} with magically implemented methods %(like \lstinline{Ref[T]})
is provided to the user as a parameter to the main method at the beginning of the execution, similarly to \lstinline{Env} in Pony.
There is a strong similarity between \Q@System@ and \Q@Ref[T]@: both are normal non magic traits. The user can define their own traits implementing \Q@System@ or \Q@Ref[T]@ as normal.
Additionally, instances of both are produced by a magic factory method producing an instance whose behaviour could not be expressed by user code.
While the factory of \Q@Ref[T]@ is publicly visible,
the factory of \Q@System@ is inaccessible to the user.

With object capabilities, we can now write "Hello, World!" in Fearless:
\begin{lstlisting}
HelloW:Main{sys -> sys.println "Hello, World!"}
Main{ .main(sys: mut System): Void } //Main as declared in standard library

\end{lstlisting}

In the same way in Java execution starts from any class with a main method, in Fearless it can start from any trait (with no abstract methods) transitively implementing \lstinline{Main}.

% This allows for abstraction over the entry-point, for example a unit test could look like this:
% \begin{lstlisting}
% MyTests:UnitTest{logger->...}
% \end{lstlisting}
% Where \lstinline{UnitTest} inherits from \lstinline{Main} and implements the main method, creates a logger and so on, leaving a single method abstract that is called from the implemented \lstinline{.main}.

\subsection{Sweetening Local State}
\label{sec:sweetening}
%While our let-in syntax works well (c.f. \autoref{sec:letIn}), we think it is inconvenient that the name of the variable ends up separated from the initialisation expression. 
R-Fearless offers an extra layer of syntactic sugar, called the `= sugar', allowing local variable declarations to be integrated into fluent APIs. In code using fluent interfaces we often have an initial receiver and then a bunch of method calls, for example, in Java Streams, we could have:
\begin{lstlisting}[language=Java]
myList.stream().map(x->x*2).filter(y->y>20).toList()//Java code
\end{lstlisting}
Here \lstinline{myList} is the initial receiver of the sequence of method calls. With our proposed sugar, a method call with the form: `\lstinline@e.m(e1, {x,self -> self ... })@', where `\lstinline{...}'
 is a sequence of method calls using \lstinline{self} as the initial receiver, can be written with a more compact and flattened syntax:
`\Q@e.m x = e1. ...@'.
This sugar allows us to define many useful methods and DSLs that simplify writing Fearless code. For example, we have a \lstinline{Block[T]} trait that has a fluent API which emulates a statement based programming language. We can obtain a \lstinline{Block[T]} by calling \lstinline{Block#} without any arguments. Using \lstinline{Block[T]}, we can now write code like this:
\begin{lstlisting}[deletekeywords={read}]
MyApp:Main{sys -> Block# //with the '= sugar'
  .var[mut Fs] fs = {sys.fs}
  .var content = {fs.read("data.txt")}
  .return {fs.write("data.txt", "Hello World "+content)}
}
MyApp:Main{sys -> Block#//same code without the '= sugar'
  .var[mut Fs]({sys.fs}, {fs, self1 -> self1
    .var({fs.read("data.txt")}, {content, self2 -> self2
    })})}
\end{lstlisting}

% As we can see above, a lot of control-flow and Fearless code inherently relies on nesting object literals.

\noindent The line `\Q@.var[mut Fs] fs = {sys.fs}@'
also shows a literal implementing a single no-args abstract method.
As in Smalltalk and other languages, in this case we can omit the \Q@->@ symbol too.

We believe that the fluent interface pattern is a great match with Fearless, and in general with expression based languages \cite{10.1145/3567512.3567533}. Usually declaring local variables would break the chain, and/or require nesting. Our sugar enables local variables inside fluent-style code.
Our sugar can be seen as a very simplified encoding of continuation passing style \cite{10.1145/3564719.3568691}.
In addition to \lstinline{.var}, \lstinline{Block[T]} offers many other methods relying on the = sugar, notably, \lstinline{.ref}, declaring a \lstinline{mut Ref[T]} of the type returned by the object literal parameter.
The Fearless implementation of \lstinline{Block[T]} is purely library code, with no magic.
As you can see from the former example, \Q@Block@, \Q@Streams@, and similar fluent APIs whose methods often take a single lambda are easier to write in Fearless because they are parsed as left associative binary operators.

Of course, the Fearless compiler could optimise code by compiling specific coding patterns in a specialised way. We would expect Blocks to be compiled away as if they were conventional statements. Numbers, booleans, and strings can also be compiled into optimised forms, and optionals may be compiled into possibly null pointers.

\section{Equational reasoning is not lost}
\label{sec:reasoning}
Thanks to RC and OC, every expression that does not use externally declared \lstinline{mut}/\lstinline{iso} references is deterministic and does not mutate any existing state.\footnote{Expressions using \lstinline{iso} can only be evaluated one time because \lstinline{iso} is affine; we are hesitant to call this `determinism', but some equational reasoning holds in this case, too. However, OC can be \lstinline{iso}, so an expression using an \lstinline{iso} may have side effects.}
That is, the outer program does not need to know or care about mutable state created inside of such expressions, and it is guaranteed that no visible side effects will happen inside this code. A library can declare such functional parts by simply declaring \lstinline{imm} or \lstinline{read} methods that do not take any \lstinline{mut}/\lstinline{iso} arguments.

It's crucial that all OC methods exhibiting non-deterministic behaviour are \mut methods.
Therefore, if an object literal captures or passes around an object capability as \lstinline{read}, it would be harmless~\cite{10.1145/1455770.1455793}.
It is thanks to this property that RC/OC languages can be extended to support 
correct caching,
representation invariants and
unobservable parallelism
\cite{gordonRefCap,42Invariants,servetto2013balloon}.

Different languages may introduce different constructs supporting those features.
Instead of extending the language syntax, one R-Fearless extension could add an \Q@IsoPod[T]@ type for storing \Q@iso@ references, with a magic factory; similar to \Q@Ref@.
A \Q@mut IsoPod[T]@ can only be created with an \Q@iso T@. It encapsulates and imprisons the isolated state, allowing it to be observed and mutated exclusively by passing closures. 
In R-Fearless it is easy to limit how closures can interact with the outer world.
Consider the definition \Q@  RF[A,R]:{ read #(a: A): R }@.\\*A method taking a \Q@rf:read RF[mut A,imm R]@
can call \Q@rf#v@ certain that such execution will not mutate any object unreachable from \Q@v@, and that the result is going to be disjoint from the MROG of \Q@v@.
If the method took a \Q@rf:read RF[read A,imm R]@, then no mutation at all would be observable from the execution of \Q@rf#v@.
If we used \Q@  F[A,R]:{ #(a: read A): imm R }@ we would have even more guarantees: the method \Q@RF#@ can capture read aliases to externally visible mutable objects, while method \Q@F#@ can only capture the immutable ones. Thus calling
\Q@f#v@ with a structurally identical \Q@v@ would always produce a structurally identical result.

For example (assuming \Q@g: mut IsoPod[Graph]@ and \Q@n: iso Node@)

\begin{lstlisting}
g.look{readG -> readG.size}
g.mutate(n, {mutG, mutN -> mutG.add(mutN)})
\end{lstlisting}

Here method \Q@.look@ takes
a \Q@read RF[read Graph,imm Num]@ and \Q@.mutate@ takes
\Q@imm F[mut Graph,mut Node,imm Void]@.
The \Q@read@ closure, captures a readonly view of the outer world and produce only immutable results.
Mutate can also take additional isolated references and they will all be (consumed and) seen as extra \mut parameters.
Mutate takes an \Q@imm@ closure.
If a \Q@read@ closure were accepted, the outer \Q@mut Isopod[Graph]@ could be captured and its mutation could be observed via a \read alias from within the \Q@.mutate@ execution.
In our supplementary material we explain better the consequence of this API design decision.
\begin{lstlisting}
g.cachedLook{ readG -> readG.countPartitions }
g.addInvariant{ readG -> readG.size < 100 }
\end{lstlisting}
Caching is now very easy to add.
We could have a \Q@.cachedLook@ method taking an immutable closure \Q@imm F[read Graph,imm Bool]@, capturing only an \imm view of the world. The system can then choose (as an unobservable optimisation) to cache the result of any \Q@.cachedLook@ and to clear the cache after any call to mutate.
Invariants can be seen as a kind of caches that must be instantly computed/recomputed, and fail with an error in case they do not hold.
This design is equivalent to L42's invariants \cite{42Invariants}, but since L42 does not have closures, they had to use an ad-hoc annotation system instead.

Parallelism can be achieved by
asynchronous variants of the look and mutate methods, returning a \Q@Future[R]@ and taking \Q@imm F@ lambdas instead of \Q@read RF@ ones.
Multiple \Q@.asyncLook@ calls can be executed in parallel, while \Q@.asyncMutate@ calls would have to be queued and executed in the submitted order.
Simple fork-join parallelism can then be obtained by the following code.
It would not be observable if the system chooses to run the \Q@.asyncMutate@ calls in parallel or sequentially.
In this example, the calls \Q@res1#@ and \Q@res2#@ are waiting for the associated computations to terminate:
\begin{lstlisting}
.forkJoin(g1: mut IsoPod[Graph], g2: mut IsoPod[Graph]): Num-> Block#
  .var[Future[Num]] res1 = {g1.asyncMutate{mutG-> mutG.partition}}
  .var[Future[Num]] res2 = {g2.asyncMutate{mutG-> mutG.partition}}
  .return {res1#+res2#}
\end{lstlisting}

The methods \Q@.asyncLook@ and \Q@.asyncMutate@ give us a solid base to implement libraries like parallel streams/iterators that could keep the parallelism unobservable to the programmer.
Numerous other primitive and unobservable forms of parallelisation enabled by reference capabilities can be considered, like reactive programming \cite{10.1145/581690.581695,10.1145/3611096.3611098}; our ongoing research is focused on identifying those most compatible with the overarching Fearless ecosystem.
% \section{Flows and Automatic Parallelism}

% \subsection{Managing mutable and immutable state}

% \subsection{Examples}

\section{Related Work}
\label{sec:related}
Featherweight Java has been a mainstay in the world of PL formalisation since it was first published, as shown by its numerous extensions \cite{bierman2003mj,Murawski2014,toplas42,10.1145/3611096.3611098}. %Fearless is both as minimal as FJ and usable in practical programming.

%Fearless could be associated with prototype based OO languages, having the same syntax for object literals and top level declarations. 

In a world without
fields, object identities and monkey patching, the distinction between prototype-based\cite {prototypes}, class-based OO, multiple inheritance and traits composition blurs. Fearless top-level declarations can be seen as both super types/classes and global singleton objects; indeed, they can be used as such if they have no abstract methods.
In our current formalism we loosely follow Sch{\"a}rli et al.'s \cite{traits} original traits model and Java interfaces with default methods, which have been shown to be able to encode Sch{\"a}rli's traits \cite{10.1145/2647508.2647520,bono2008traits,bettini2018java}.
Overall, we are confident that our model avoids the object-based pitfalls described in \cite{jones2016object}.

Fearless can be seen as an interface-based programming language (as introduced by Wang et al. \cite{classlessJava}), where interface-based programming enables multiple inheritance by decoupling state and behaviour. However, unlike Wang et al.'s proposed `Classless Java', Fearless does not make use of fields or need any code generation for `abstract state operations'. Additionally, Fearless is more minimal by avoiding the need for static methods. $\lambda$-based OO programming \cite{servetto2021lambda} is the closest to a Fearless-like OO model, but they rely on lambdas implementing a single abstract method.
They discuss how privateness is directly supported in $\mathrm{FJ}_\lambda$ (and thus also in Fearless) thanks to the absence of dynamic casts: just expose the object through an interface with fewer methods. They also show that super-calls are easy to encode.
In their future work, they wonder what could be a good way to integrate imperative state updates and imperative coding styles. In our work, we have proposed a solution using RC/OC and our \Q{Block} library.

Other work with RC/OC \cite{dalaCaps,ponyCaps}  uses explicit language syntax when working with \lstinline{iso} and/or promotions/recovery. Our design follows M\#'s \cite{gordonRefCap} and L42's \cite{42Recovery} approach to consume \iso and promote the expression transparently. Fearless limits those promotions to method calls. Having two ways to type-check a method call may cause a naive implementation to run in exponential time. Our prototype implementation uses some simple, well-known strategies \cite{10.1145/604131.604133} to avoid having to try all options; we omit a detailed discussion on this topic for space reasons.

Capturing state in a flexible and predictable way is a big concern for Fearless, needing closures for even simple control flow like if-else.
There is no unified approach for capturing RC or affine types into lambdas/closures.
Pony offers explicit syntax for how state is captured, but this can get quite verbose; a default behaviour is also provided.
Rust uses move semantics, which requires the closure to take ownership of the object and invoking the closure moves that object, preventing multiple accesses (\Q{FnOnce}) \cite{rustClosures}.

In many famous type systems without RC
\cite{10.1145/286936.286947,ostlund2008ownership,10.1145/2663171.2663188,regions,hogg1991islands} aliasing control is obtained by tracking objects for their whole lifetime and propagating this knowledge across the program. Instead, reference capabilities allow objects to be born untracked (\Q{mut}), and information can be recovered from the usage context.

In R-Fearless we used object capabilities \cite{10.1145/361932.361937,objCapsE} to support side-effects like IO without breaking equational reasoning. Either the Fearless Heart or R-Fearless could be extended to support algebraic effects, like those seen in Call$\mathcal{E}$ \cite{10.1145/3359591.3359731} instead of object capabilities.

There is a wealth of topics that our work connects to, and we cannot do justice to them all. Some of these topics include: the push towards encapsulating fields in OO code \cite{meyer1988eiffel,effectiveJava}; the extensive literature around reference capabilities \cite{10.1007/3-540-45337-7_2,javari,giannini2016types,giannini2019flexible,nobleRefCapsRegions} and their related concepts of ownership \cite{10.1145/286936.286947}, regions \cite{regions}, islands \cite{hogg1991islands} and uniqueness \cite{externalUniqueness};  object capabilities, both in relation with RC \cite{42Invariants,10.1145/3611096.3611098} and as an independent concept \cite{10.1145/361932.361937,objCapsE,Drossopoulou2020}; unobservable caching \cite{10.1145/604131.604133}, invariants \cite{invariants} and their related concepts of runtime verification \cite{Meredith2011} and static verification \cite{staticVerification}; automatic parallelisation as a compiler optimisation \cite{friedman1978aspects,Harris2007,Trilla2015,ertel2019towards};  the benefits of functional programming and equational reasoning \cite{10.1145/24697.24706,Gibbons2011}; OO design patterns with a focus on visitors, which are foundational for our pattern matching \cite{gangOfFour}; fluent interfaces \cite{fluentInterface}, continuation-passing style \cite{Appel1989} and their relations to domain specific languages \cite{10.1145/3567512.3567533}.

\section{Conclusions and Future Work}
\label{sec:conclusion}
While FJ represents OO programming as it was popularly understood in the 90s by Java, C\#, and C++, we believe that Fearless better represents what OO programming is now, where fluent interfaces and lambdas are everywhere, realising the early vision of untyped Smalltalk/Ruby but in a nominally typed setting.

As future work, many Fearless extensions are possible\footnote{We discuss a few possible extensions like those in more detail in our supplementary material.} including exceptions, better sub-typing for generic containers, more precise typing for \read containers, method type refinements, more fine grained reference capabilities, 
sealed hierarchies, private classes and packages.

\printbibliography

% \newpage
\appendix
\input{appendix}

\end{document}

%% file: formalism.tex
The formal model disregards all sugar (like optional parentheses), assumes that all the inference has already been applied and that arity based overloading has been elided as discussed earlier.

The design decision to have all types living at top level, including the ones declared inside of a method body, greatly simplifies the formalism. It allows for a single implicit top level trait table to be shared across all the notations, like what happens in the FJ \cite{fj} formalisation, where a fixed top-level class table is globally available to all rules.

Because $\D\OS\Ts\CS$ is of form $T$, we often use \NT{T} for the list of implemented traits to lighten the notation.
We use an underscore ($\_$) when a part of a term is irrelevant, which is often the case for the list of implemented traits.

\subsubsection{Well formedness}
We now list the syntactic well formedness requirements. They are all standard except for the last two, which we will discuss in the detail.

\begin{itemize}
    \item All declaration names, appearing anywhere, are disjoint.
    \item In an object literal, all method names are disjoint.
    \item In all \NT{sig}, all \Xs and \NT{\overline{x}} are disjoint.
    \item There is no shadowing of \NT{X} or \NT{x}.
    \item Inheritance relations are acyclic.
    \item $\sName$ in top-level declarations must be \lstinline{'this}.
    \item The \D of \L declared inside method bodies are not implemented by any other \L. In Java's terminology, they are `final types'.
    \item An \L declared inside of a method body have no free type variables. This means that the \Xs in \L must contain all type variables used in such \L.
    \item The \Xs of an \L declared inside of a method body must only contain type variables that are currently in scope.
\end{itemize}

%  I really would like this discussion to happen outside of the formalism
% ---
To better understand how an \L inside of a method body works, consider our \Q{Opt[T]}, where we instead give a name to the inner literal, thus declaring a new visible top level type.
\begin{lstlisting}
Opt:{ #[T](t: T): Opt[T]  -> {m -> m.some(t)} }//former version
Opt:{ #[T](t: T): Some[T] -> Some[T]:Opt[T]{m -> m.some(t)} }//alternative

\end{lstlisting}

The type `\Q@Some[T]@' can exist both inside and outside the method body (the return type of \Q@#@). While many alternative designs were possible, Fearless makes the \Q@[T]@ type parameter explicit. Thus, the declaration of \Q@Some@ is instantiated using `\Q@T@', and redeclares `\Q@T@' as a generic type parameter simultaneously. We call this mechanism \emph{funnelling generic type parameters}.
% ----

\subsubsection{Reduction}
Our reduction rules are conventional. Our only values are object literal terms of the form \L.
\textsc{call-top} invokes a method from the top-level traits table, which happens if the method is not directly defined in the object literal that the method is being called on. Our notation for extracting the method signature with instantiated generics and body `$\mathrm{meths(T})$' is defined later in the type system. We use standard substitution notation to bind our variables in $e$. The `$\mathrm{dom(\Ms)}$' notation extracts all method names from a sequence of methods.
%The $e$ of the method body may be an alpha renamed version of the expression present in the top level method.
%This may be needed because some $L$ inside $e$ may declare some $X$ or $x$ that would shadow those currently in scope. 
\textsc{call-lit} invokes the method directly on the object literal. Finally, we have two contextual rules. Because Fearless's syntax is so minimal, we can avoid inserting a dedicated evaluation context into the grammar and can simply define two \textsc{ctx} rules.
 
\metarule{call-top}{
    \L  = \D\OS\Ts'\CS\QQ{:} \_\  \OC\sName\ \Ms\ \CC\qquad
    \m \notin \mathrm{dom}(\Ms)\\
    \m \OS\Xs\CS\OR\x_1\QQ{:}\T_1, \ldots, \x_n\QQ{:}\T_n\CR\col \T \ \ \point  \e \com \in \mathrm{meths}(\D\OS\Ts'\CS)
}{
    \L\, \m \OS\Ts\CS\OR\Ls\CR
    \rightarrow
     \e[\Xs=\Ts][\QQ{this}, \x_1, \ldots,  \x_n=\L,\Ls]
}{}\hspace{-0.4cm}
\metarule{ctx-recv}{
     \e  \rightarrow \e'
}{
     \e\,  \m \OS\Ts\CS\OR\es\CR
    \rightarrow
    \e'\,\m \OS\Ts\CS\OR\es\CR
}{}\\
\metarule{call-lit}{
    \L  = \D\OS\_\CS\QQ{:} \_\ \OC\sName\ \Ms\CC\qquad
    \m \in \mathrm{dom}(\Ms)\\
     \m \OS\Xs\CS\OR\x_1\QQ{:}\T_1, \ldots, \x_n\QQ{:}\T_n\CR\col \T \ \ \point  \e \com \in \Ms
}{
    \L\, \m \OS\Ts\CS\OR\Ls\CR
    \rightarrow
    \e[\Xs=\Ts][\x , \x_1, \ldots,  \x_n=\L,\Ls]
}{}
\metarule{ctx-arg}{
     \e  \rightarrow \e'
}{
    \L\,\m \OS\Ts\CS\OR\Ls \com  \e  \com \es\CR
    \rightarrow
    \L\, \m \OS\Ts\CS\OR\Ls \com \e' \com \es\CR
}{}
% \\[5ex]

As you can see, Fearless follows the straightforward top to bottom, left to right call by value execution strategy.
We omit the formal definition for the standard replacement operations $\e[\Xs=\Ts]$ and $\e[\xs=\Ls]$. However, we included this definition in our mechanisation of this model, provided in our supplementary material.
\subsubsection{Trait Type system}
The type system relies on a conventional local variable environment \begin{Grammar}
    \Production{\Gamma}{\NT{x}_1:\NT{T}_1 \ldots \NT{x}_n:\NT{T}_n}{}
\end{Grammar}

The rule \textsc{all-ok} type-checks a program by extracting the set of all \Ls (including those within expressions) from all of the top-level declarations ($\L_1 \ldots \L_n$).

\metarule{all-ok}{
    \ \Ls = \mathrm{allLs}(\L_1) \ldots \mathrm{allLs}(\L_n)\\
    \forall \L \in \L_1\ldots\L_n, \ \emptyset\vdash^{\Ls}\L : \mathrm{OK}
}{
    \L_1 \ldots \L_n : OK
}{}
\begin{minipage}[r]{0.4\textwidth}
    \begin{Define}{27}{0}{\mathrm{allLs}(\NT{e}) = \Ls,\quad \mathrm{allLs}(\M) = \Ls}
\Line{\mathrm{allLs}(\L)
  }{
  \L,\    \mathrm{allLs}(\M_1) \ldots \mathrm{allLs}(\M_n)
  }{}
\With{
    \text{where}\ \L =  \D\OS\Xs\CS\QQ{:} \_\  \{\NT{\sName}\ \M_1 \ldots \M_n \}
}{}

\Line{\mathrm{allLs}(\NT{sig}\com)}{\emptyset}{}

\Line{\mathrm{allLs}(\m\OS\Xs\CS\OR\_\CR\col\T\ \point \e\com)}{\mathrm{allLs}(\e)}{}

\Line{\mathrm{allLs}(\x)}{\emptyset}{}
\Line{\mathrm{allLs}(\e_0\m\OS\Xs\CS\OR\e_1, \ldots \e_n\CR)}{
    \mathrm{allLs}(\e_0) \ldots \mathrm{allLs}(\e_n)
}{}
\end{Define}
\end{minipage}

Notation `$\mathrm{allLs}$' is defined by recursively extracting every literal from expressions and methods.
We explicitly pass
\Ls to rule \textsc{Lit-ok} using syntax `$\vdash^{\Ls}$', to represent that this is the top level trait table that will be implicitly available to rule
\textsc{Lit-ok} and to any other rule and notation called by it.

% A \NT{Top} is well-typed if every method inside it is well-typed, and $\mathrm{meths(\D\OS\Xs\CS)}$ is defined. We check that methods are well-typed with \textsc{abs-ok} and \textsc{impl-ok} for abstract and implemented methods respectively.

A \L is well-typed by \textsc{lit-ok} if all its methods are well-typed and `overrideOk' and `implementOk' hold. \textsc{abs-ok} states that all abstract methods are well-typed. \textsc{impl-ok} states that an implemented method is well-typed if its body can be typed to match the method's return type. We support subsumption, so a sub-type of the declared return type is also valid. Note how, in the type context for checking the body, we compose the trait's type variables with the method's type variables and compose all visible variables (including the self-name) with the method parameters.

\metarule{Lit-ok}{
    \Xs, \Gamma\,\x:\D\OS\Xs\CS \vdash \M_1 : \mathrm{OK}\ 
    \ \ldots\ 
    \Xs, \Gamma
    \,\x:\D\OS\Xs\CS
    \vdash \M_n : \mathrm{OK}\\
    \mathrm{overrideOk}(\D\OS\Xs\CS)\qquad
    \mathrm{implementOk}(\D\OS\Xs\CS)
    %\mathrm{meths}(\D\OS\Xs\CS)
}{
    \Gamma \vdash \D\OS\Xs\QQ{]:} \_\  \{\NT{\sName}\ \NT{M}_1 \ldots \NT{M}_n \} : \mathrm{OK}
}{}
% \begin{minipage}[l]{0.75\textwidth}
% \end{minipage}
\begin{minipage}[r]{0.3\textwidth}
\metarule{abs-ok}{}{
    \Xs, \Gamma \vdash \NT{sig}\com : \mathrm{OK}
}{}
\metarule{impl-ok}{
    \Xs\ \Xs',\ \Gamma\ \Gamma' \vdash \NT{e} : \NT{T}
}{
    \Xs, \Gamma \vdash \NT{m}\OS\Xs'\CS\OR\Gamma'\CR\col\NT{T}\ \point \NT{e}\com : \mathrm{OK}
}{}
\end{minipage}

Our notations `$\mathrm{overrideOk}$' and `$\mathrm{implementOk}$' do the heavy lifting of checking that methods from implemented traits are compatible. They rely on new syntax, where \begin{Grammar}
    \Production{DM}{\D\OS\Ts\CS\fstop\M}{}
\end{Grammar}\!\!\!\!\!\!
 represents a method belonging to a specific trait. 
 \begin{Grammar}
    \Production{mtype}{\m\OS\Xs\CS\col \T_1 \ldots \T_n \rightarrow \T}{}
\end{Grammar}
 \!\!\!\!\! represents a method type, and notations $\mathrm{mtype(\DMs)}$ and $\mathrm{mtype}(M)$ extract the \mtype{}s from the \DMs or \M. We use the notations $\mathrm{name}(\NT{mtype})$ and \(\mathrm{name}(M)\) trivially extract the method name \m. 

Notation `$\mathrm{dmeths}$' is used inside $\mathrm{overrideOk}$ and $\mathrm{implementOk}$.
As shown below, it collects all the \DM pairs from all the transitively implemented traits. Since the implements relation is acyclic, this process terminates. Note how the trait's generic type variables are replaced during collection.

\begin{Define}{5}{0}{\mathrm{dmeths}(D\OS\Ts\CS) = \DMs\ \DMs'}
%    \Line{\mathrm{dmeths}(D\OS\Ts\CS)}{\DMs\ \DMs'}{where}
    \With{
        D\OS\Xs\CS\col \T_1 \ldots \T_n\ \OC\sName\ \M_1 \ldots \M_k \CC \in \Ls
    }{}
    \With{\DMs = D\OS\Ts\CS.\M_1[\Xs=\Ts]\ \ldots\ D\OS\Ts\CS.\M_k[\Xs=\Ts]}{}
    \With{\DMs' = \mathrm{dmeths}(T_1[\Xs=\Ts]) \ldots \mathrm{dmeths}(T_n[\Xs=\Ts])}{}
\end{Define}

Our notation `$\mathrm{overrideOk}$' checks that all method signatures of equally named methods are compatible. $\mtype_1 \simeq \mtype_2$ holds if there is a way to alpha-rename the \Xs of the two {\mtype}s into free (type) variables so that the two {\mtype}s are syntactically identical. That is, in this minimal formalisation, we require the same exact signature. Simplicity is the only reason for this design decision; an alternative design could support return type refinement or co/contra variant implementations like many other languages.

\begin{Define}{5}{0}{\mathrm{overrideOk}(T)\ \text{holds iff }\ \forall \mtype_1, \mtype_2 \in \mathrm{mtype}(\mathrm{dmeths}(T)),}
    \With{\mathrm{if}\ \mathrm{name}(\mtype_1) = \mathrm{name}(\mtype_2)\ \mathrm{then }\ \mtype_1 \simeq \mtype_2}{}
\end{Define}

Our notation `implementOk' checks that no conflicting implementations exist for the same method.
That is, our traits handle multiple inheritance and conflicts similarly to
Sch{\"a}rli et al.'s original traits model~\cite{traits},
where the order of implementation does not matter and the programmer must specify an explicit override in case of conflicts.

\begin{Define}{5}{0}{\mathrm{implementOk}(T)\ \text{holds iff }\ \forall \DM_1, \DM_2 \in \mathrm{dmeths}(T),
\mathrm{conflict}(\DM_1,\DM_2)\ \mathrm{then}}
    \With{
        \exists \DM_3 \in \mathrm{dmeths}(T),\ \mathrm{alternative}(\DM_3, \DM_1)
        \ \mathrm{and}\ \DM_3 \leq \DM_1\ \mathrm{and}\ \DM_3 \leq \DM_2
    }{}
\end{Define}

The formal definition defined above relies on `conflict' and `alternative' and states that for all conflicting pairs $\DM_1$ $\DM_2$, an alternative $\DM_3$ that beats them exists. Note that we do not require alternative$(\DM_3,\DM_2)$, thus $\DM_2$ could be selected as the $\DM_3$ solving the conflict.
\begin{Define}{0}{0}{
    \mathrm{alternative}(\DM_1, \DM_2)\qquad
    \mathrm{conflict}(\DM_1, \DM_2)\qquad
    \DM_1 \leq \DM_2
}
    \Line{
        \mathrm{alternative}(T_1.M_1, T_2.M_2)
    }{T_1 \neq T_2\ \mathrm{and}\ \mathrm{name}(M_1) = \mathrm{name}(M_2)}{}

    \Line{\mathrm{conflict}(\DM_1, \DM_2)}{
        \mathrm{alternative}(\DM_1, \DM_2),\  
        \mathrm{not}\ \mathrm{abs}(\DM_2)\
        \mathrm{and}\
        \mathrm{not}\ \DM_1 \leq \DM_2
    }{}

    \Line{T_1.M_1 \leq T_2.M_2}{T_1 \leq T_2}{}
\end{Define}

$\DM_1$ and $\DM_2$ are in `alternative' if they have the same name but originate from different traits. $\DM_1$ and $\DM_2$ are in `conflict' if they are alternatives; the second is not abstract, and the first does not beat the second.
Note how conflicts are directional. Because `implementOk' checks all pairs, every pair will be checked for conflicts in both directions. A \DM beats another one (notation $\DM_1 \leq \DM_2$) if the originating trait of the first  (transitively) implements the originating trait of the second one.
Finally, we can define the set of all methods of a type \T:

\begin{Define}{5}{0}{M \in \mathrm{meths}(T)\ \mathrm{iff}}
    \With{D\OS\Ts\CS.M \in \mathrm{dmeths}(T)
    \text{ and }\ \forall \DM \in \mathrm{dmeths}(T),\ \mathrm{not}\ \mathrm{conflict}(D\OS\Ts\CS.M, \DM)}{}
\end{Define}

The idea behind `meths' is that there could be many methods with the same name in $\mathrm{dmeths}(T)$, and we select the one that beats them all. Such method must exist because `implementOk' holds.

\subsubsection{Expression Type system}
The only novel expression type judgement is \textsc{lit-t}; 
Here we check that we correctly funnel the type variables and we ensure that all methods are not abstract. This could be because they are implemented in \Ms or because we inherit an implementation.
Then we call \textsc{lit-ok}.

\metarule{lit-t}{
    \Xs \subseteq \Xs'\qquad
    \forall \NT{M} \in \mathrm{meths}(\D\OS\Xs\CS), \ \text{not abs}(\M)\\
    \L=\D\OS\Xs\CS\QQ{:} \_ \ \OC\sName\ \Ms\CC\qquad
    \Gamma\vdash
    \L : \mathrm{OK}
}{
    \Xs',\ \Gamma \vdash \L : \D\OS\Xs\CS
}{}

\textsc{subs-t}, \textsc{var-t}, and \textsc{call-t} are all conventional.
The \Xs of the method extracted from `meth' may have been alpha renamed in order to not shadow the \Xs' in scope.
 Our sub-typing is also conventional. Note how the rule \textsc{lit-sub} embodies also reflexivity.

\metarule{subs-t}{
    \Xs, \Gamma \vdash \NT{e} : \NT{T}' \qquad
    \NT{T}' \leq \NT{T}
}{
    \Xs, \Gamma \vdash \NT{e} : \NT{T}
}{}
\metarule{var-t}{}{
    \Xs, \Gamma \vdash \NT{x} : \Gamma(\NT{x})
}{}
\metarule{lit-sub}{
 \D\OS\Xs\QQ{]:} \NT{T}_1, \ldots, \NT{T}_n\, \OC\sName\ \_\CC \in \Ls\\
    \NT{T} \in \D\OS\Ts\CS, \NT{T}_1[\Xs = \Ts], \ldots, \NT{T}_n[\Xs = \Ts]
}{
    \D\OS\Ts\CS \leq \NT{T}
}{}
\metarule{trn-sub}{
    \NT{T}_1 \leq \NT{T}_2\qquad \NT{T}_2 \leq \NT{T}_3
}{
    \NT{T}_1 \leq \NT{T}_3
}{}
\metarule{call-t}{
    \Xs', \Gamma \vdash \e_0 : \T_0 \qquad
    \m\OS\Xs\CS\OR\x_1\QQ{:}\T_1, \ldots, \x_n\QQ{:}\T_n\CR\col\T\ \ \point \e\com \in \mathrm{meths}(\T_0)\\
    \Xs', \Gamma \vdash \e_1 : \T_1[\Xs = \Ts]\  \ldots\  \Xs', \Gamma \vdash \e_n : \T_n[\Xs = \Ts]\qquad \Xs'\text{disj}\, \Xs
}{
    \Xs', \Gamma \vdash \e_0\,\m\OS\Ts\CS\OR\e_1,\ldots\e_n\CR : \T[\Xs = \Ts]
}{}

We have mechanised this formal model's type system and its related notation with PLT-Redex \cite{redex}.
We type-checked many code examples, to help reassure us that our formal model behaves as expected.
% Our mechanisation is available as supplementary material. We additionally have written 95 test cases to help verify that important components in our model are working correctly.

%% file: appendix.tex
% \documentclass[acmsmall,screen,review,anonymous,nonacm,natbib=false]{acmart}

% \input{preamble}

% \RequirePackage[
%   datamodel=acmdatamodel,
%   % style=acmauthoryear,
%   style=numeric,
%   ]{biblatex}

% %% Declare bibliography sources (one \addbibresource command per source)
% \addbibresource{fearless-journey.bib}

% \title{The Fearless Journey: Supplementary Material}

% \begin{document}
% \maketitle

\newpage
\section{Prototype Compiler}
We have developed a prototype compiler over the past fourteen months, with more than 22,000 lines of Java code and more than 700 commits. Our compiler implements Fearless with a number of extensions like overloading on method $R$s, sealed traits, packages, type aliases, return type refinement in method selection, adapter-based sub-typing for generics, and bounded reference capabilities for generic parameters. The compiler also supports hygienic reference capabilities. We will discuss some of these extensions later in this document. We are currently compiling to Java source code, using interfaces to represent our top-level trait declarations and anonymous inner classes to represent object literals in method bodies. We have partial support for exceptions and \Q@IsoPod@s.

Our standard library uses fluent interfaces extensively. For example, this working and running code sample uses our `\Q@Block@' library and our `\Q@Iter@' library, which both have a fluent API:

\begin{lstlisting}
package test
alias base.Int as Int, alias base.Str as Str,
alias base.List as List, alias base.Block as Block,
alias base.caps.FIO as FIO,

IterFind:base.Main{sys -> Block#
    .var l1 = {List#(35, 52, 84, 14)}
    .assert{l1.iter
      .map{n -> n * 10}
      .find{n -> n == 140}
      .isSome}
    .var msg = {l1.iter
      .filter{n -> n < 40}
      .flatMap{n -> List#(n, n, n).iter}
      .map{n -> n * 10}
      .str({n -> n.str}, ",")}
    .var io = {FIO#sys}
    .return {io.println(msg)}
    // prints 350,350,350,140,140,140
}
\end{lstlisting}

\section{PLT-Redex Mechanisation of The Fearless Heart}
PLT-Redex \cite{redex} is a domain-specific language for mechanising formal models, specifically of programming languages. Notably, it can generate terms based on the defined language's reduction rules and type system for test-case generation. This tool enables us to verify the semantics of our formal model in a more 1:1 manner than our prototype compiler.

The Fearless Heart's formalisation in PLT-Redex is a little over eight hundred lines long with test cases including the \Q{Opt[T]} example from section 2 in our paper. We also have developed a tool for turning Fearless programs from our prototype compiler into terms for our PLT-Redex model, enabling us to quickly type-check and run substantive chunks of Fearless source code with our model.

We have a working mechanisation of the Fearless Heart's type system, as presented in this paper, and a partial implementation of our reduction semantics. We also present an auxiliary model that is closer to our prototype compiler's implementation of Fearless and offers full reduction semantics and type-checking for the core language without generics.

In figure~1 you can see the type system's derivations for checking that \textsc{all-ok} holds, and in figure~2 you can see the reduction of a simple method call.

\begin{figure}[h]
  \label{fig:derivations}
  \includegraphics[width=0.9\linewidth]{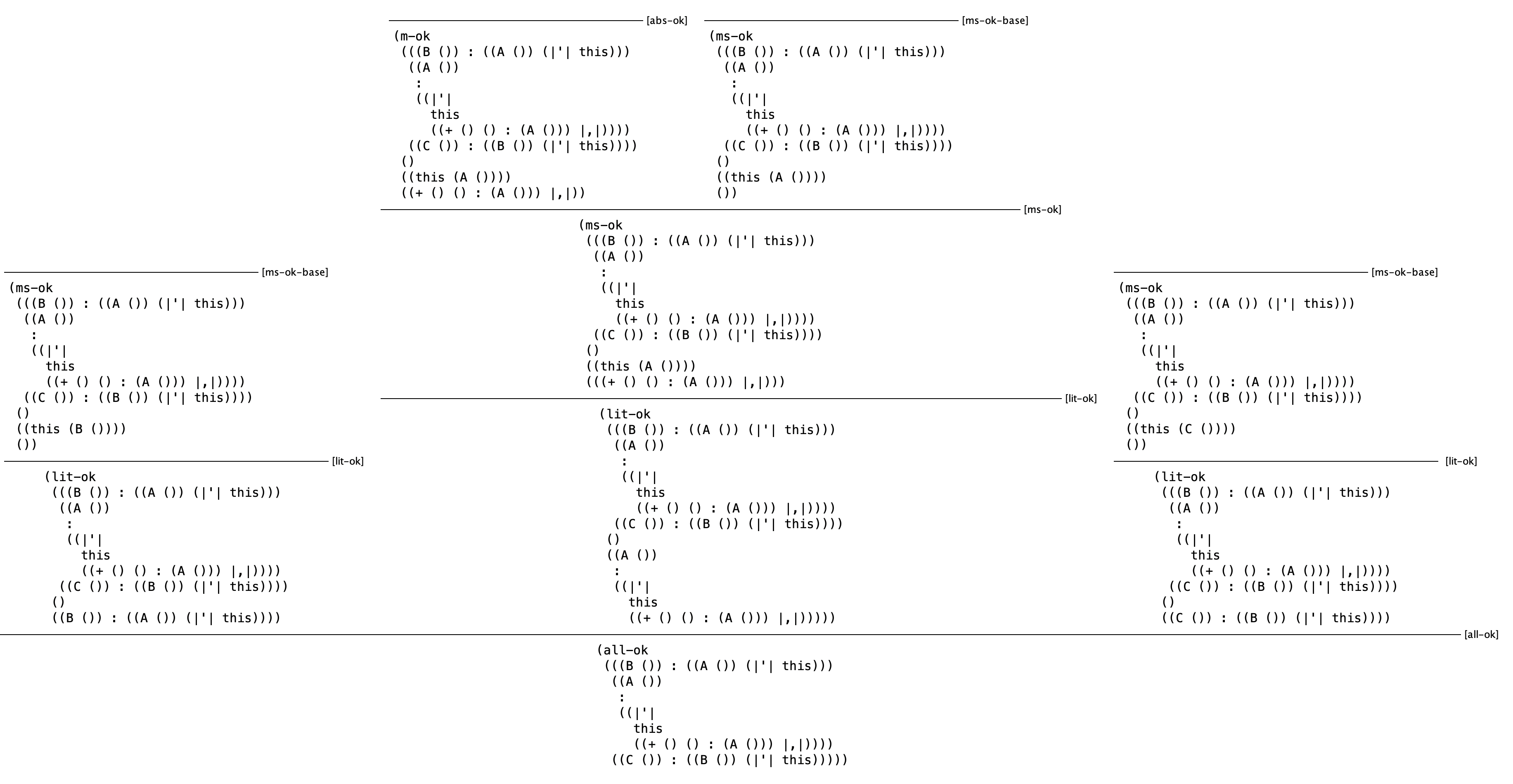}
  \caption{An example derivation from type-checking a Fearless Heart program with our PLT-Redex model.}
  \Description{A diagram showing how the model is proving all-ok for a simple program with three literals, where one of them has an abstract method. The pairs are (from left to right) all-ok -> lit-ok -> ms-ok-base, all-ok -> lit-ok -> ms-ok -> (abs-ok, ms-ok-base), all-ok -> lit-ok -> ms-ok-base}
\end{figure}

\begin{figure}[h]
  \label{fig:reduction}
  \includegraphics[width=0.9\linewidth]{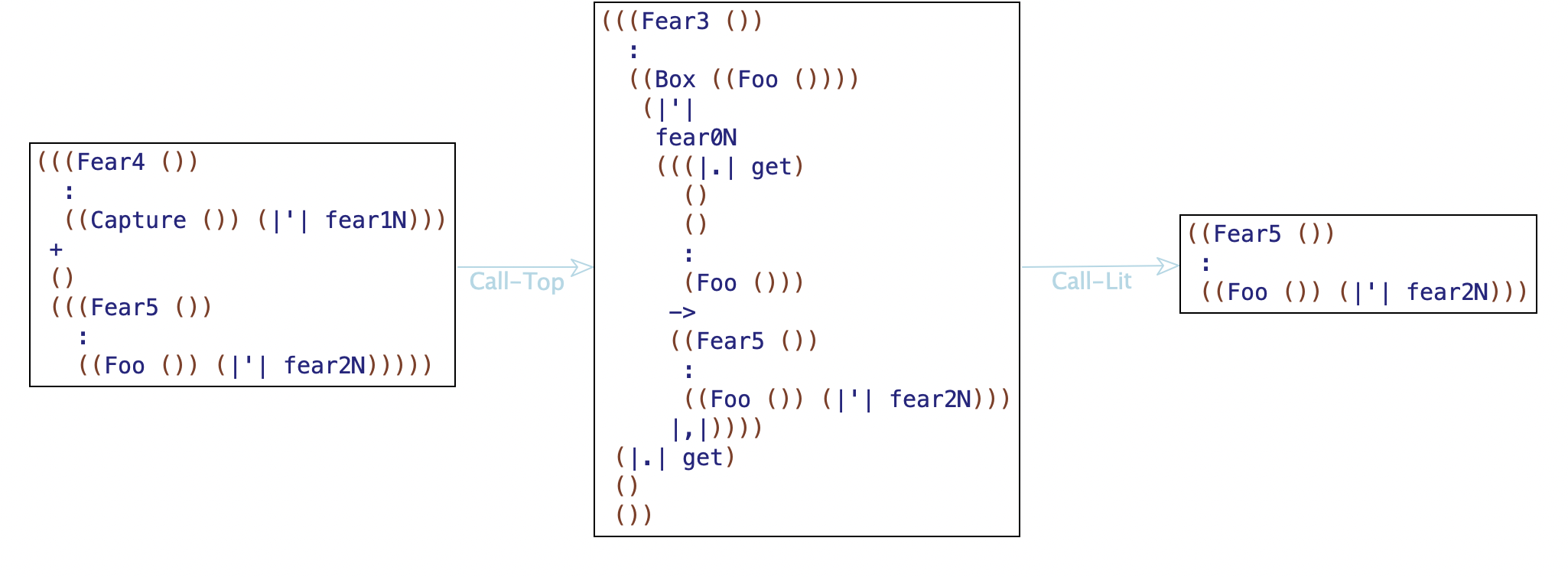}
  \caption{An example reduction our Fearless Heart PLT-Redex model of a program that captures a variable in a container and then extracts it.}
  \Description{A three-step reduction from a method call (Call-Top) to the method call that extracts our variable from an object literal that captures it (Call-Lit) to the captured value.}
\end{figure}

\section{A concrete example of invariants and caching}

In this section, we present a concrete code example showing how invariants may work in Fearless.
This code is highly inspired by the first example of \cite{42Invariants}, to show the similarities of our approaches.

 Consider the iconic Range data type, offering getters and setters for min/max values. We expect a representation invariant that the min value must always be smaller than the max value. We may also want to cache the result of the string method returning the current string representation of the current range object.

 Let us divide the task into two steps. First, we define a \lstinline{MinMax} type offering min/max but no invariant or cache. Basically a mutable Nat*Nat tuple:
\begin{lstlisting}
FMinMax:F[Nat,Nat,MinMax]{min, max -> Block#
  .ref _min = {min}
  .ref _max = {max}
  .return{MinMax{
    read .min:Nat-> _min.get, 
    read .max:Nat-> _max.get,
    mut  .min(n:Nat):Void-> _min.set(n),
    mut  .max(n:Nat):Void-> _max.set(n),
    read .toS:Str-> _min.get+","+_max.get  
}}}
\end{lstlisting}
Note how we defined the \lstinline{MinMax} type internally. In this way, the \lstinline{MinMax} trait is final, and we can only create an object literal that implements it by calling \lstinline{FMinMax#}. Then, we can define the desired \lstinline{Range} type using an \lstinline{FRange} factory:
\begin{lstlisting}
FRange:F[Nat,Nat,Range]{(min,max -> Block#
  .isoPod minMax= { MinMax#(min,max) } .invariant { r-> r.min < r.max }
  .return {Range{
    read .min:Nat->minMax.look{r->r.min}, 
    read .max:Nat->minMax.look{r->r.max},
    mut  .min(n:Nat):Void->minMax.mutate{r->r.min(n)},
    mut  .max(n:Nat):Void->minMax.mutate{r->r.max(n)},
    mut  .set(min:Nat, max:Nat):Void->
      minMax.mutate{r->Block#(r.min(min),r.max(max),Void)}
    read .toS:Str->minMax.cachedLook{r-> "Range["+r.toS+"]"},
}}}
\end{lstlisting}
In the previous example, we imagine that Block is extended with a \Q@.isoPod@ method behaving like \Q@.var@ or \Q@.ref@, but returning an object asking for an invariant before returning the control to the block.

Now any observable object of type Range will have a valid invariant. The method \Q@.set(min,max)@ is quite interesting.
The invariant is checked only after both fields have been updated, and there is a moment where one field only has been updated. In this moment the invariant of the outer \Q@Range@ object may not hold.
Depending on the specific details of the \Q@.mutate@ API this may or may not be a problem.
If \Q@.mutate@ takes a immutable \Q@F@ lambda, then
this is not a problem because the outer \Q@Range@ object is not reachable form inside the lambda: it must be \mut to accept a call to \Q@.mutate@, and a \mut object can not be captured by the immutable lambda.
On the other side, if \Q@.mutate@ takes a \read \Q@RF@ lambda, then the outer \Q@Range@ object may be visible inside of it, and we may observe a broken invariant.
This is why we \Q@.look@ and \Q@.mutate@ take different kinds of closures.
%used our representation invariant library (\lstinline{Repr}). The library roughly looks like this:
% \begin{lstlisting}
% Repr{ #[T](t:iso T):mut Repr[T]-> ... } // omitted
% Repr[T]:{
%   read .look[R](f: read RF[read T, R]): R,
%   read .mutate[R](f: read RF[mut T, R]): R,
%   mut .invariant(f: F[read T, Bool]): Void,
%   mut .cached[R](f: F[read T, R]): mut CachedProperty[R],
% }
% RF[A,R]:{ read #(a: A):R }
% CachedProperty[R]:{ mut #:R }
% \end{lstlisting}

% \begin{itemize}
%     \item \lstinline{Repr#} creates an instance of \lstinline{Repr[T]} by hiding an initial \lstinline{iso T} as a \lstinline{mut T} captured reference.
%     \item \lstinline{Repr[T].look/1} takes a \lstinline{read RF} object literal. By being \lstinline{read} and taking a \lstinline{read} parameter, the execution of this \lstinline{RF} object literal is guaranteed not to mutate any visible reference. \lstinline{Repr[T].look/1} calls \lstinline{f} on the hidden \lstinline{mut T} reference and returns the result.
% \end{itemize}
% TODO: maybe the rest of this but it's all kinda wrong

\section{R-Fearless Formal Model}
In the paper, we described the R-Fearless formal model by showing the interesting differences over the Fearless model. Here, 
for clarity, we show a self-contained model of the R-Fearless type system.
We omit the reduction since it would follow Featherweight Java's standard extension to support state.
\subsection{Grammar and Well formedness}

\begin{Grammar}
    \Production{L}{\D\OS\Xs\QQ{]:} \NT{D_1}\OS\Ts_1\CS, \ldots, \D_n\OS\Ts_n\CS \{\NT{\sName}\ \Ms\}}{}\\
    \Production{M}{\NT{sig\com} \mid \NT{sig}\quad \point\ \NT{e\com}}{}\\
    \Production{e}{
        \NT{x}
        \mid \NT{e}\fstop\NT{m}\OS\Ts\CS\OR\NT{\overline{e}}\QQ{)}
        \mid \NT{R}\ \L
    }{}\\
    \Production{sig}{\NT{R}\ \NT{m}\OS\Xs\CS\OR\NT{x_1\QQ{:}T_1, \ldots, x_n\QQ{:}T_n}\CR\col\NT{T}}{}\\
    \Production{T}{
        \NT{R}\ \NT{D\OS\overline{T}\CS}
        \mid \NT{X} \mid
        \NT{R}\ \NT{X}}{}\\
    \Production{R}{\QQ{imm} \mid \QQ{iso} \mid \QQ{read} \mid \QQ{mut}}{}\\
    \Production{DM}{\D\OS\Ts\CS\fstop\M}{}\\
    \Production{mtype}{\m\OS\Xs\CS\col \T_1 \ldots \T_n \rightarrow \T}{}\\
    \Production{\Gamma}{\NT{x}_1:\NT{T}_1 \ldots \NT{x}_n:\NT{T}_n}{}\\
    %\Production{\NT{Top}}{\Gamma \vdash \L}{}
\end{Grammar}
\begin{itemize}
    \item All declaration names appearing anywhere are disjoint.
    \item In an object literal, all method names are disjoint.
    \item In all \NT{sig}, all \Xs and \NT{\overline{x}} are disjoint.
    \item There is no shadowing of \NT{X} or \NT{x}.
    \item Inheritance relations are acyclic.
    \item $\sName$ in top-level declarations must be \lstinline{'this}.
    \item The \D of \L declared inside method bodies can not be implemented. In Java's terminology, they are `final types'.
    \item An \L declared inside of a method body have no free type variables. This means that the \Xs in \L must contain all type variables used in such \L.
    \item The \Xs of an \L declared inside a method body must only contain type variables currently in scope.
    \item  A
The parameter of type $\iso\,\_$ can be used only once in the method body or any number of times but
only inside object literals. 
\item In a valid $\D\OS\Ts\CS$ any $\T \in \Ts$ is not of form $\iso\,\_$.
\end{itemize}
\subsection{Type system}

\begin{Define}{10}{0}{\mathrm{allLs}(\NT{e}) = \Ls,\quad \mathrm{allLs}(\M) = \Ls}
\Line{\mathrm{allLs}(\L)
  }{
  \L,\    \mathrm{allLs}(\M_1) \ldots \mathrm{allLs}(\M_n)
  }{}
\With{
    \text{where}\ \L =  \D\OS\Xs\CS\QQ{:} \_\  \{\NT{\sName}\ \M_1 \ldots \M_n \}
}{}

\Line{\mathrm{allLs}(\NT{sig}\com)}{\emptyset}{}

\Line{\mathrm{allLs}(\m\OS\Xs\CS\OR\_\CR\col\T\ \point \e\com)}{\mathrm{allLs}(\e)}{}

\Line{\mathrm{allLs}(\x)}{\emptyset}{}
\Line{\mathrm{allLs}(\e_0\m\OS\Xs\CS\OR\e_1, \ldots \e_n\CR)}{
    \mathrm{allLs}(\e_0) \ldots \mathrm{allLs}(\e_n)
}{}
\end{Define}

\metarule{all-ok}{
    \ \Ls = \mathrm{allLs}(\L_1) \ldots \mathrm{allLs}(\L_n)\qquad
    \forall \L \in \L_1\ldots\L_n, \ \emptyset\vdash^{\Ls}\mut\,\L : \mathrm{OK}
}{
    \L_1 \ldots \L_n : OK
}{}

\metarule{R-Lit-ok}{
    \Gamma_i = (\Gamma, \x:\R_0\,\D\OS\Xs\CS)[\R_0,\mathrm{rcOf}(\M_i)]\qquad
    \Xs, \Gamma_1 \vdash \M_1 : \mathrm{OK}\ 
    \ \ldots\ 
    \Xs, \Gamma_n\vdash \M_n : \mathrm{OK}\\
    \mathrm{overrideOk}(\D\OS\Xs\CS)\qquad
    \mathrm{implementOk}(\D\OS\Xs\CS)
    %\mathrm{meths}(\D\OS\Xs\CS)
}{
    \Gamma \vdash \R_0\ \D\OS\Xs\QQ{]:} \_\  \{\NT{\sName}\ \NT{M}_1 \ldots \NT{M}_n \} : \mathrm{OK}
}{}

\begin{Define}{24}{0}{
    \Gamma[R_0,R] = \Gamma'\qquad \T[\R]=\T'
}
%Case1
  \Line{(\x: \T, \Gamma)[\R_0,\R_1]
  }{
  \x: \T[\imm],\ \   \Gamma[\R_0,\R_1]
  \quad \text{ with }  \T = \iso \_ \text{ or } \imm \_
  }{}
%Case2
  \Line{(\x: \T, \Gamma)[\R_0,\R_1]
  }{
  \x: \T,\ \   \Gamma[\R_0,\R_1]
  \quad \text{ with }  \R_0\in\{\iso,\mut\}
  \text{ and } \R_1\in\{\iso,\mut\}
  }{}
%Case3
  \Line{(\x: \T, \Gamma)[\R_0,\imm]
  }{
  \x: \T[\imm],\ \   \Gamma[\R_0,\imm]
  \quad \text{ with }  \R_0\in\{\iso,\mut,\read\}
  }{}
%Case4
\Line{(\x: \T, \Gamma)[\R_0,\read]
  }{
  \x: \T[\read],\ \   \Gamma[\R_0,\read]
  \quad \text{ with }  \R_0\in\{\iso,\mut,\read\}
  }{}
%CaseEmpty
\Line{\emptyset[\R_0,\R_1]
  }{\emptyset
  }{}
%T[R]
\With{
\R\, \D[\Ts][\R'] = \R'\D[\Ts]\qquad
\R\, \X[\R'] = \R'\X\qquad
\X[\R'] = \R'\X
}{}
\end{Define}

\metarule{abs-ok}{}{
    \Xs, \Gamma \vdash \NT{sig}\com : \mathrm{OK}
}{}
\metarule{impl-ok}{
    \Xs\ \Xs',\ \Gamma\ \Gamma' \vdash \NT{e} : \NT{T}
}{
    \Xs, \Gamma \vdash \R\,\NT{m}\OS\Xs'\CS\OR\Gamma'\CR\col\NT{T}\ \point \NT{e}\com : \mathrm{OK}
}{}

Note how in rule \textsc{impl-ok}, the receiver capability \R is correctly ignored since the receiver is already contained in $\Gamma$. That is, the \Q@this@ object is captured by the method instead of being a parameter of it.

\begin{Define}{5}{0}{\mathrm{dmeths}(D\OS\Ts\CS) = \DMs\ \DMs'}
    \With{
        \D\OS\Xs\CS\col \D_1\OS\Ts_1\CS \ldots \D_n\OS\Ts_n\CS\ \OC\sName\ \M_1 \ldots \M_k \CC \in \Ls
    }{}
    \With{\DMs = D\OS\Ts\CS.\M_1[\Xs=\Ts]\ \ldots\ D\OS\Ts\CS.\M_k[\Xs=\Ts]}{}
    \With{\DMs' = \mathrm{dmeths}(\D_1\OS\Ts_1\CS[\Xs=\Ts]) \ldots \mathrm{dmeths}(\D_1\OS\Ts_1\CS[\Xs=\Ts])}{}
\end{Define}

\begin{Define}{5}{0}{\mathrm{overrideOk}(T)\ \text{holds iff }\ \forall \mtype_1, \mtype_2 \in \mathrm{mtype}(\mathrm{dmeths}(T)),}
    \With{\mathrm{if}\ \mathrm{name}(\mtype_1) = \mathrm{name}(\mtype_2)\quad \mathrm{then }\ \mtype_1 \simeq \mtype_2}{}
\end{Define}

\begin{Define}{5}{0}{\mathrm{implementOk}(T)\ \text{holds iff }\ \forall \DM_1, \DM_2 \in \mathrm{dmeths}(T),
\mathrm{conflict}(\DM_1,\DM_2)\ \mathrm{then}}
    \With{
        \exists \DM_3 \in \mathrm{dmeths}(T),\ \mathrm{alternative}(\DM_3, \DM_1)
        \ \mathrm{and}\ \DM_3 \leq \DM_1\ \mathrm{and}\ \DM_3 \leq \DM_2
    }{}
\end{Define}

\begin{Define}{0}{0}{
    \mathrm{alternative}(\DM_1, \DM_2)\qquad
    \mathrm{conflict}(\DM_1, \DM_2)\qquad
    \DM_1 \leq \DM_2
}
    \Line{
        \mathrm{alternative}(\D_1\OS\Ts_1\CS.\M_1, \D_2\OS\Ts_2\CS.\M_2)
    }{\D_1\OS\Ts_1\CS \neq \D_2\OS\Ts_2\CS\ \mathrm{and}\ \mathrm{name}(\M_1) = \mathrm{name}(\M_2)}{}

    \Line{\mathrm{conflict}(\DM_1, \DM_2)}{
        \mathrm{alternative}(\DM_1, \DM_2)\ 
        \mathrm{not}\ \mathrm{abs}(\DM_2),\ \mathrm{not}\ \DM_1 \leq \DM_2
    }{}

    \Line{\D_1\OS\Ts_1\CS.\M_1 \leq \D_2\OS\Ts_2\CS.\M_2}{\mut\,\D_1\OS\Ts_1\CS  \leq \mut\,\D_2\OS\Ts_2\CS}{}
\end{Define}

\begin{Define}{5}{0}{M \in \mathrm{meths}(T)\ \mathrm{iff}}
    \With{D\OS\Ts\CS.M \in \mathrm{dmeths}(T)
    \text{ and }\ \forall \DM \in \mathrm{dmeths}(T),\ \mathrm{not}\ \mathrm{conflict}(D\OS\Ts\CS.M, \DM)}{}
\end{Define}

\metarule{subs-t}{
    \Xs, \Gamma \vdash \NT{e} : \NT{T}' \qquad
    \NT{T}' \leq \NT{T}
}{
    \Xs, \Gamma \vdash \NT{e} : \NT{T}
}{}
\metarule{var-t}{}{
    \Xs, \Gamma \vdash \NT{x} : \Gamma(\NT{x})
}{}
\\
\metarule{call-t}{
    \Xs', \Gamma \vdash \e_0 : \D\OS\Ts_0\CS \qquad \R_0 \leq \R\qquad \Xs'\text{disj}\, \Xs\\
    \R\,\m\OS\Xs\CS\OR\x_1\QQ{:}\T_1, \ldots, \x_n\QQ{:}\T_n\CR\col\T\ \ \point \e\com \in \mathrm{meths}(\R_0\D\OS\Ts_0\CS)\\
    \Xs', \Gamma \vdash \e_1 : \T_1[\Xs = \Ts]\ \ldots\  \Xs', \Gamma \vdash \e_n : \T_n[\Xs = \Ts]
}{
    \Xs', \Gamma \vdash \e_0\,\m\OS\Ts\CS\OR\e_1,\ldots\e_n\CR : \T[\Xs = \Ts]
}{}
\\
\metarule{prom-call-t}{
    \Xs', \Gamma \vdash \e_0 : \R_0\, \D\OS\Ts_0\CS\qquad \R_0 \in \{\imm,\iso\} \qquad \R_0 \leq \R
    \qquad \Xs'\text{disj}\, \Xs\\
    
    \R\ \m\OS\Xs\CS\OR\x_1\QQ{:}\T_1, \ldots, \x_n\QQ{:}\T_n\CR\col\T\ \ \point \e\com \in \mathrm{meths}(\R_0\D\OS\Ts_0\CS)\\
    \Xs', \Gamma \vdash \e_1 : \PromP{\T_1[\Xs = \Ts]} \ \ldots\  \Xs', \Gamma \vdash \e_n : \PromP{\T_n[\Xs = \Ts]}
}{
    \Xs', \Gamma \vdash \e_0\m\OS\Ts\CS\OR\e_1,\ldots\e_n\CR : 
    \PromR{\T[\Xs = \Ts]}
}{}\\[3ex]
\\
\metarule{R-lit-t}{
    \Xs \subseteq \Xs'\qquad
    \forall \NT{M} \in \mathrm{meths}(\D\OS\Xs\CS), \ \text{if callable}(\R,\M) \text{ then not abs}(\M)\\
    \L=\R\ \D\OS\Xs\CS\QQ{:} \_ \ \OC\sName\ \Ms\CC\qquad
    \forall \M \in \Ms, \ \text{callable}(\R,\M)\qquad
    \Gamma\vdash
    \L : \mathrm{OK}
}{
    \Xs',\ \Gamma \vdash \R\ \L : \R\ \D\OS\Xs\CS
}{}
\\
\metarule{trn-sub}{
    \NT{T}_1 \leq \NT{T}_2\\ \NT{T}_2 \leq \NT{T}_3
}{
    \NT{T}_1 \leq \NT{T}_3
}{}
\metarule{R-sub}{
    \R \leq \R'
}{
    \T[\R]\leq \T[R']
}{}
\metarule{lit-sub}{
 \D\OS\Xs\CS\QQ{:} \D_1\OS\Ts_1\CS, \ldots, \D_n\OS\Ts_n\CS\, \OC\sName\ \_\,\CC \in \Ls\\
   \T_i=\R\,\D_i\OS\Ts_i\CS[\Xs = \Ts]\\
    \T \in \R\,\D\OS\Ts\CS, \T_1, \ldots, \T_n
}{
    \R\,\D\OS\Ts\CS \leq \T
}{}

\begin{Define}{0}{0}{
    \mathrm{callable}(\R, \M) \text{ iff }\R\in\{\QQ{imm},\QQ{read}\} \text{ implies  rcOf}(\M)\not\in\{\QQ{mut},\QQ{iso}\}
}
\end{Define}

Sub-typing for $R$ is defined as below:
\begin{Define}{0}{0}{
    % R \leq R' = 
    \QQ{iso} \leq \_ \leq \QQ{read}
}
\end{Define}

\begin{Define}{8}{0}{
    \Prom\R = \R'\qquad
    \PromP\T = \T'\qquad
    \PromR\T = \T'\qquad
}
  \With{
    \Prom\mut = \iso\qquad
    \Prom\read = \imm\qquad
    \Prom\R = \R\ \mathrm{where}\ \R \notin \{ \mut, \read \}
  }{}

\With{
    \PromP{\R\, \D\OS\Ts\CS} = \Prom\R\ \D\OS\Ts\CS\qquad
    \PromP{\R\, \X} = \Prom\R\ \X\qquad
    \PromP\X = \iso\, X
}{}

\With{
    \PromR{\R\, \D\OS\Ts\CS} = \Prom\R\ \D\OS\Ts\CS\qquad
    \,\,\PromR{\R\, \X} = \Prom\R\ \X\qquad
    \PromR\X = \imm\,\X
}{}
\end{Define}

Note on reductions:
The type system, as presented, can only type-check source programs.
To type-check an R-Fearless program in the middle of the execution, we would need to type the current memory and give a type of locations.
Plus, we would need to relax the syntax of object literals \L to allow for any kind of \T in the \D\OS\Xs\CS.

 \section{Better subtyping for generic containers}
 %Extending sub-typing with the adapter pattern
Generic variance is supported by many languages, in Java with wildcards, in Scala with \Q@+/-@ and in C\# with \Q@in/out@.
Fearless has only simple generics without generic bounds and/or variance.
We would like to keep our generics simple, but we would like more subtyping when possible; we realised that in many cases, the user could simply write a wrapper object.
Consider for example, an imperative List type:
\begin{lstlisting}
List[T]:{
  mut  .get(i: Num): T,
  read .rget(i: Num): read T,
  read .size,
  mut .add(e: T),
  mut .set(i: Num, e: T),
  }
Person:{..}
Student:Person{..}
\end{lstlisting}
Clearly, a \Q@ps: mut List[Person]@ is not a subtype or a supertype of a \Q@ss: mut List[Student]@.

\Q@ps.get(i)@ would return persons that may not be students, and \Q@ss.add(bob)@ can not be called if \Q@bob@ is not a student.
However, a \Q@imm List[Person]@ is logically a supertype of a \Q@imm List[Student]@.
A determined programmer could even manually implement a wrapper:
\begin{lstlisting}
Wrapper:{ #(ss: List[Student]): List[Person] -> {
  .rget(i)->ss.rget(i),
  .size->ss.size,  
}}
\end{lstlisting}
With the code above, \Q@Wrapper#(ss)@ can be used anywhere a List[Person] is needed. Since Fearless does not offer dynamic casts, there is no significant difference between subtyping and wrapping.
As discussed in the RC section, the set of available methods and methods \Q@.add,.set,.get@ are not implemented by an \imm object literal.
We are considering expanding the R-Fearless sub-typing so that the two types are recognised as subtypes any time such a wrapper would be possible.

We plan to add a rule \RuleName{adapt-sub-t}, that will check if all the methods callable on a subtype candidate could be identically called on a supertype candidate.
We forecast to encounter problems in cases like the iconic \Q@List[T].concat(List[T]):List[T]@ method:
Does it satisfy the `adapt rule'?
A simple-minded metarule
would attempt to generate an infinite proof, and an implementation following it would go in stack overflow.

Our `adapt rule' would also allow for an \Q@imm List[mut Person]@ to be passed where an \Q@imm List[imm Person]@ is required. Those two types are identical since the whole ROG of an immutable object is immutable, and the adapt rule would recognise this. 
We wonder if we should apply the adapt rule to a few white-listed cases, to generic types with the same head (e.g. two Lists with different generic arguments), or to allow it for any kind of types, basically opening up to a restricted form of structural typing.

\section{More precise typing for read containers}
In our paper, we discussed how \Q@.rget@ is not satisfactory in the case of read collections.
Both Pony and M\# use an algebra of reference capabilities or types, expressing syntactically and explicitly viewpoint adaptations. This feature allows for types whose RC changes depending on the RC of other types.
Those features allow for complex types that we would rather avoid.
We have considered many options, and our current best candidate is to add another form to our types:
\begin{Grammar}
    \Production{\T}{\R\,\D\OS\Ts\CS\mid\X\mid\R\,\X\mid \QQ{readImm}\, \X}{}
\end{Grammar}
a \Q[morekeywords={readImm}]@readImm X@ would be \imm if \X is instantiated with an \imm, and \read in all the other cases.
Then, we could declare
\Q[morekeywords={readImm}]@read .rget:readImm T@ and solve this current limitation.
We are trying to understand if there is some other case where we would need access to the full generality that a type algebra offers. If we can not find one, we will consider formally model \Q[morekeywords={readImm}]@readImm@ as a future Fearless extension.
\Q[morekeywords={readImm}]@readImm@ would behave as a subtype of \read and a supertype of \imm, and would require an extension for notations $\PromP{\_}$ and $\PromR{\_}$ used during promotion.

\section{Let-in}
Even without our sugar for local variables, The Fearless Heart already has all the tools to encode local variables via let-in. We can encode it either as a function with two arguments:
\begin{lstlisting}
Let:{ #[T,R](x: T, f: F[T,R]): R -> f#x }
... Let#(12 + foo, {x -> x*3}) ... // usage

\end{lstlisting}
Or as a curried function, allowing us to write the \lstinline{.in} syntax explicitly:
\begin{lstlisting}
Let:{ #[T](x: T): In[T] -> {f -> f#x } }
In[T]:{ .in(f: F[T,R]): R }
... Let#(12 + foo) .in {x -> x*3} ... // usage
\end{lstlisting}

\section{Exceptions}
For an RC and OC system to work, we must consider exceptions carefully. 
This is a well-known situation; see \cite{42Invariants} for a detailed discussion about exception safety. In short, R-Fearless can be extended with two kinds of exceptions: unchecked exceptions, which a magic \lstinline{Try.catch} method that can capture with strong exception safety \cite{10.1007/3-540-39953-4_6,10.1145/1924520.1924523}, and non-deterministic exceptions, like a stack overflow, that can only be captured by a method of a dedicated capability object. Overall, using exceptions is quite rare in Fearless, with sum types that can represent failure cases being more common.
This extension is partially implemented in our prototype.

% , method type refinements, hygienic types,

\section{Sealed hierarchies, Private classes and
Packages}
Fearless allows for final types but not for sealed subtype hierarchies.
This is unfortunate since sealed types allow us to model ADT much more closely and can be used to perform better compile time optimisations.
We have extended the Fearless prototype with a special \Q@Sealed@ interface and with a concept of packages. Packages would contain top-level declarations, pretty much as they do in Java packages.
\Q@Sealed@ would have no runtime semantic but would influence the type system: a trait \Q@Foo@ implementing \Q@Sealed@ could only be implemented either in its own package or with an \L of form \Q@FreshName[]:Foo[@\Ts\Q@]{}@.
This allows for \Q@Bool@, \Q@Num@, \Q@Str@, \Q@Opt@, \Q@Void@, and many others to be sealed, unlocking many compiler optimisations while allowing for object literal \Q@False@/\Q@True@/\Q@Opt[Person]@ to be used in any package.
Packages also allow for package-private traits.
While (as discussed in our related work)  private methods are intrinsically supported in Fearless, privateness at the trait level can be introduced as an extension. In our prototype compiler, traits whose name starts with an underscore can only be used inside their own package.

%We could extend Fearless by allowing object literals defined inside methods to be extended inside of the same method.
%However this would prevent our clean encoding for booleans.
%Every time we write \Q@True@ we are actually defining a new anonymous trait extending \Q@True@.
%If \Q@True@ was defined inside of a method, it could capture some local method parameter, and thus could not possibly be extended outside of his method.
%In this extension, booleans may be encoded as follows:
%\begin{lstlisting}
%True:{#}
%\end{lstlisting}
%We would force the user to write something like %\Q@True#@

\section{Hygienic modifiers}
\newcommand\readH{\QQ{readHyg}\xspace}
\newcommand\mutH{\QQ{mutHyg}\xspace}

RC promotions are great, but they can only apply when only immutable and isolated state is taken in input. Former work on language L42\cite{42Recovery} introduced more modifiers to handle promotions able to take in input mutable state that was guaranteed not to be involved in the promotion.
Here, we outline an R-Fearless extension that could serve a similar role. We add two modifiers: \readH and \mutH.
They are hygienic in the sense that they follow this property: \emph{If the MROG of a hygienic references is disjoint from the MROG of any other reference  in a given moment in time, it will stay disjoint for the rest of the execution. As an exception, the data of isolated references can be injected into hygienic references.}
However, hygienic references do not enforce the MROG to start as disjoint. Such a property can easily be derived, for example, by initialising hygienic references with \iso references.
Moreover, since newly created \mut object literals are clearly not in the MROG of any pre-existing hygienic reference, they will stay disjoint.

When can a hygienic reference be captured?
Again, we are trying to focus on simplicity and usability, so we have simple capturing rules. The object literals will simply see a $\Gamma$ where all the hygienic references have been removed.

However, hygienic references come with a new promotion that we call `scoped-promotions'.
By the scoped promotion, a method has an alternative signature, where
zero or one \mut is replaced with \mutH,
all the other \mut are replaced with \iso, and all the \readH with \read.
In a scoped promotion, the method result is `hygienised': if \mut or \read it becomes \mutH or \readH.

%New1, not discussed. Useful if methods takes readonly.. but this is discouraged
%sig[mut=iso, read=imm, readonly=imm]
The good news is that the other promotion stays the same:
All the \mut are turned in \iso, and all the \read
parameters are turned in \imm.
This means that \mutH and \readH are untouched, and a method taking those as formal arguments can take actual arguments of those types during promotion.
This allows to produce isolated data-structures while accessing and mutating externally visible mutable objects.
Consider the following scanner example, inspired by a similar example present in the \Q@lent@ RC literature:
\begin{lstlisting}[morekeywords={mutHyg,readHyg}]
    Parse:{
      .parse(s:mutHyg Scanner):mut AST->..
      .parseAll(s:mut Scanner):iso Ast->this.parse(s)
    }
\end{lstlisting}
The method \Q@.parseAll@ takes a \Q@mut Scanner@, but passes it as \Q[morekeywords={mutHyg,readHyg}]@mutHyg@ to \Q@.parse@. Since the \Q[morekeywords={mutHyg,readHyg}]@mutHyg@ MROG can not mix with freshly created \mut results, the method result can be promoted to \iso.

With the two promotions together, a \mut into \mut method can be used both as an \iso into \iso and as a \Q[morekeywords={mutHyg,readHyg}]@mutHyg@ into \Q[morekeywords={mutHyg,readHyg}]@mutHyg@.

Any method taking just a bunch or \read parameters and producing a \read result can be used as a method taking a bunch of \Q[morekeywords={mutHyg,readHyg}]@readHyg@ parameters and returning a \Q[morekeywords={mutHyg,readHyg}]@readHyg@ result.

An object literal created inside a method called under a scoped promotion can transparently capture references that are hygienic outside of that method scope. 
The new object could have captured any amount of (externally) \Q[morekeywords={mutHyg,readHyg}]@readHyg@ references, and if returned as \Q[morekeywords={mutHyg,readHyg}]@mutHyg@, it could have also captured a single \Q[morekeywords={mutHyg,readHyg}]@mutHyg@ reference too.

We think that the above explanation could be understood by a reader expert with L42 style \Q@lent@ refererence capabilities, but would be insufficient for someone that have not seen this flavour of RC yet.
We leave a deep dive and a clear, exhaustive explanation of all the details about hygienic references to future work. We suspect that \Q[morekeywords={mutHyg,readHyg}]@mutHyg@ and \Q[morekeywords={mutHyg,readHyg}]@readHyg@ as presented here are more expressive than the ones in the lent RC literature. We plan to examine and justify this claim in future work.

% \section{Reference Capability bounds}

% \section{Method body promotions}

% \section{Flows}

% \printbibliography

% \end{document}